\documentclass[preprint]{elsarticle}

\usepackage{color}
\usepackage{graphicx}% Include figure files
\usepackage{dcolumn}% Align table columns on decimal point
\usepackage{bm}% bold math
\usepackage{hyperref}% add hypertext capabilities

\usepackage{epstopdf}
\usepackage{amsmath}

\usepackage{subfig}
\usepackage{multirow}

\usepackage{pgf}
\newcommand{\ket}[1]{\ensuremath{\left| #1 \left\rangle \right. \right.}}

\newcommand{\elemm}[3]{\ensuremath{\left\langle #1 \left| #2 \right| #3 \right\rangle}}
\newcommand{\prodbk}[2]{\ensuremath{\left\langle #1 \left| #2 \right\rangle \right.}}

\journal{Computer Physics Communications}

\begin{document}

\title{Efficient On-the-Fly Interpolation Technique for Bethe-Salpeter 
Calculations of Optical Spectra}% Force line breaks with \\

\author[a]{Yannick Gillet\corref{author}}
\author[a]{Matteo Giantomassi}
\author[a]{Xavier Gonze}

\cortext[author] {Corresponding author.\\\textit{E-mail address:} yannick.gillet@uclouvain.be}
\address[a]{ European Theoretical Spectroscopy Facility, 
Institute of Condensed Matter and Nanosciences, 
Universit\'e catholique de Louvain, 
Chemin des \'etoiles 8, bte L7.03.01, 
1348 Louvain-la-Neuve, Belgium}

\date{\today}% It is always \today, today,

\begin{abstract}
The Bethe-Salpeter formalism represents the most accurate method available 
nowadays for computing neutral excitation energies and optical spectra of 
crystalline systems from first principles. 
Bethe-Salpeter calculations yield very good agreement with experiment but are 
notoriously difficult to converge with respect to the sampling of the 
electronic wavevectors.
Well-converged spectra therefore require significant computational and memory 
resources, even by today's standards. 
These bottlenecks hinder the investigation of systems of great technological 
interest. They are also barriers to the study of derived 
quantities 
like piezoreflectance, thermoreflectance or resonant Raman intensities.

We present a new methodology that decreases the workload 
needed to reach a given accuracy. It is based on a double-grid on-the-fly 
interpolation within the Brillouin 
zone, combined with the Lanczos algorithm. It achieves 
significant speed-up and reduction of memory requirements.
The technique is benchmarked in terms of accuracy on silicon, gallium arsenide 
and lithium fluoride. 
The scaling of the performance of the method as a function of the 
Brillouin Zone point density is much better than a conventional implementation.
We also compare our method with other similar techniques proposed in the 
literature.
\end{abstract}	
\maketitle


%\tableofcontents

\section{Introduction}

The calculation of optical properties from first principles can be achieved 
with different levels of approximation and different computational requirements. 
The Bethe-Salpeter Equation (BSE) in the framework of Many-Body 
Perturbation Theory is the most precise and sophisticated approach to 
compute the macroscopic dielectric function including the attractive 
electron-hole interaction~\cite{Onida2002}. This 
formalism 
has been used since 
1998~\cite{Albrecht1998} for the first-principles computation of the optical 
spectra of semiconductors and insulators.
Even though its application to simple 
materials is reasonably frequent nowadays, the calculation of optical 
properties of complex 
materials with more than two dozen atoms in the unit cells is still 
challenging (see e.g. the work of Kresse \textit{et al.}~\cite{Kresse2012} and 
Rinke \textit{et al.}~\cite{Rinke2012}). Algorithmic 
improvements~\cite{Rohlfing2000, Paier2008, Gruning2011, 
Sander2015},
and new theoretical developments to introduce temperature 
dependence~\cite{Marini2008}, or 
to compute resonant Raman intensities from derivatives of optical 
response~\cite{Gillet2013} are active domains of research. 

Independently of the approximations used, the precise description of the 
dielectric properties usually requires a large number of wavevectors to sample 
the Brillouin Zone (BZ). Each wavevector of the BZ, indeed, gives contributions 
at different transition energies and small changes in the mesh density can
induce oscillations in the dielectric properties~\cite{Gillet2013}. 
The construction of the BSE Hamiltonian requires the computation of matrix 
elements 
connecting different points in the wavevector mesh. 
This computation is the most time-consuming part of the BSE flow and 
its cost renders well-converged results difficult to 
achieve.
The extraction of the macroscopic dielectric function from the inversion of the 
Hamiltonian matrix constitutes the last step of the BSE flow.
Two different approaches are commonly used nowadays: direct diagonalization and 
Lanczos algorithms following the seminal work of Haydock~\cite{Haydock1980}.
The Lanczos approach was employed for the first time for the solution of the 
BSE 
by Benedict \textit{et al.}~\cite{Benedict1998a,Benedict1999}. It is based 
on the construction of a  chain of vectors obtained by performing simple 
matrix-vector operations.

Since 1998, different numerical methods have been proposed to help 
reduce the computational cost.
Rohlfing and Louie~\cite{Rohlfing2000} (abbreviated RL) proposed a double-grid 
technique 
in which the kernel is interpolated starting from a 
homogeneous coarse mesh. This approach is used, for example, in the 
BerkeleyGW code~\cite{Deslippe2012}.
Another technique by Fuchs \textit{et al.}~\cite{Fuchs2008}, 
uses an inhomogeneous mesh and refines the sampling to extract specific 
information for bound 
excitons. The large Hamiltonian matrices obtained with these methods are then 
treated 
either by direct diagonalization or with the Lanczos 
algorithm.
Alternatively, one can perform the average of optical properties using
several independent shifted coarse grids, as introduced by Paier \textit{et 
al.}~\cite{Paier2008} and, later, by Gillet \textit{et al.}~\cite{Gillet2013}. 

A completely different approach to the BSE problem has been proposed recently 
by Kammerl\"ander \textit{et al.}~\cite{Kammerlander2012}.
The authors focus on a single frequency and avoid the setup of the entire 
matrix and the direct diagonalization by using an iterative technique that 
takes advantage of a double grid to solve the Dyson equation.

In the present work, we propose a new method that combines the RL 
interpolation with the Lanczos-Haydock algorithm without requiring the storage 
of the full matrix.
We also generalize the RL approach to include 
multi-linear interpolation, and we reformulate the algorithm to render it 
more scalable and less memory demanding. We present different levels of 
interpolation, with different computational loads. 

This article is organized as follows. In Section \ref{sec2}, we describe the 
main equations and the iterative approach used to solve the BSE. 
Section \ref{sec3} presents the interpolation methodology while the technical 
details of the implementation are discussed in Section \ref{sec4}. 
Finally, 
in Section \ref{sec5}, we apply our technique with different interpolation 
levels to three different crystalline systems: bulk silicon, gallium arsenide 
and lithium fluoride. We conclude with a comparison between our method and 
other similar techniques proposed by Paier \textit{et al.}~\cite{Paier2008} and 
Gillet \textit{et al.}~\cite{Gillet2013}, and the work of Kammerl\"ander 
\textit{et al.}~\cite{Kammerlander2012}.

\section{The Bethe-Salpeter Equation and the Lanczos recursion algorithm 
\label{sec2}}

In the so-called Tamm-Dancoff Approximation (TDA)~\cite{Gruning2009}, the 
matrix 
elements of the BSE Hamiltonian
in the transition space, i.e. products of valence and conduction bands, are 
given by
\begin{align}
H_{vc\boldsymbol{k},v'c'\boldsymbol{k}'} = \left( \varepsilon_{c\boldsymbol{k}} 
- \varepsilon_{v\boldsymbol{k}} \right) \delta_{\boldsymbol{k} \boldsymbol{k}'} 
\delta_{vv'} \delta_{cc'} + K_{vc\boldsymbol{k},v'c'\boldsymbol{k}'},
\label{bse1}
\end{align}
where the kernel $K$ is defined as
\begin{align}
 K_{vc\boldsymbol{k},v'c'\boldsymbol{k}'} =
 2 \elemm{vc\boldsymbol{k}}{\bar{v}}{v'c'\boldsymbol{k}'} -
   \elemm{vc\boldsymbol{k}}{W}{v'c'\boldsymbol{k}'},
\label{bse2}
\end{align}
with
\begin{align}
\elemm{vc\textbf{k}}{\bar{v}}{v'c'\textbf{k}} &= \int \int 
\psi_{v\textbf{k}}(\textbf{r}) \psi_{c\textbf{k}}^*(\textbf{r}) 
\bar{v}(\textbf{r} - \textbf{r}') \psi_{v'\textbf{k}'}^*(\textbf{r}') 
\psi_{c'\textbf{k}'}(\textbf{r}') d\textbf{r}' d\textbf{r} \label{bse3a}  \\ 
\elemm{vc\textbf{k}}{W}{v'c'\textbf{k}'} &= \int \int 
\psi_{v\textbf{k}}(\textbf{r}) \psi_{v'\textbf{k}'}^*(\textbf{r}) 
W(\textbf{r},\textbf{r}') \psi_{c\textbf{k}}^*(\textbf{r}') 
\psi_{c'\textbf{k}'}(\textbf{r}') d\textbf{r}' d\textbf{r}.
\label{bse3b}
\end{align}

In the above expressions, $v$ and $c$ stands for valence and conduction band 
indices, $\textbf{k}$ is a wavevector in the BZ, 
$\varepsilon_{n\textbf{k}}$ and $\psi_{n\textbf{k}}$ are the energies and 
wavefunctions of band $n$ at point $\textbf{k}$. Equation~\eqref{bse3a} 
represents 
the so-called exchange term and takes into account local-fields effects. 
$\bar{v}$ is 
a modified bare Coulomb potential obtained from the bare potential 
$v(\boldsymbol{G})$ by setting the $\boldsymbol{G} = 0$ component 
to zero.
The expression in Eq.\eqref{bse3b} is usually referred to as the direct 
term and takes into account the static 
screened Coulomb interaction, $W$, through the inverse dielectric function 
$\epsilon^{-1}(\textbf{r},\textbf{r}')$ via
\begin{align}
 W(\textbf{r},\textbf{r}') = \int d\textbf{r}'' 
\epsilon^{-1}(\textbf{r},\textbf{r}'') v(\textbf{r}''-\textbf{r}'), 
\label{eq:Wcoul}
\end{align}

The wavefunctions and eigenenergies are usually obtained from a 
standard Kohn-Sham calculation~\cite{Hohenberg1964,Kohn1965} and a 
scissors operator may be employed to 
mimic the opening of the gap introduced by 
the GW approximation~\cite{Onida2002}. 
For BSE applications, it is common to compute the direct term with a screened 
interaction obtained within the Random-Phase Approximation 
(RPA)~\cite{Adler1962,Wiser1963}. Alternatively, one can employ the much 
cheaper model dielectric function proposed by Cappellini 
in Ref.~\cite{Cappellini1993}.

Finally, the macroscopic dielectric function $\varepsilon_M(\omega)$ is given 
by
\begin{align}
 \varepsilon_M(\omega) = 1 - \lim_{\boldsymbol{q} \rightarrow 0} 
v(\boldsymbol{q}) \elemm{P(\boldsymbol{q})}{((\omega + i\eta) - 
H)^{-1}}{P(\boldsymbol{q})} \label{epsmacro}
\end{align}
where $v(\boldsymbol{q})$ is the Fourier transform of the Coulomb interaction, 
$P(\boldsymbol{q})$ are the oscillator matrix elements
\begin{align}
 P(\boldsymbol{q})_{vc\boldsymbol{k}} = 
\elemm{c\boldsymbol{k+q}}{e^{i\boldsymbol{q}.\boldsymbol{r}}}{v\boldsymbol{k}}
\end{align}
evaluated for small $\boldsymbol{q}$ and $\eta$ is a broadening 
factor.

The solution of the Bethe-Salpeter equation is a two-step 
process. First, the matrix elements of the Hamiltonian are computed from 
Eqs.~(\ref{bse1}-\ref{bse3b}). Then, the 
macroscopic dielectric function is derived using Eq. \eqref{epsmacro}.

In order to avoid the inversion of large matrices, Lanczos-based iterative 
techniques (called Lanczos algorithm in this work) can be used to obtain the 
macroscopic dielectric function. By using Krylov subspaces, it is possible to 
express Eq.~\eqref{epsmacro} in terms of a continued fraction formula.

The Lanczos algorithm can be summarized as follows. We start by setting
\begin{align}
 b_1 &= 0\\
 \ket{\psi_{1}} &= \frac{\ket{P(\boldsymbol{q})}}{\|\ket{P(\boldsymbol{q})}\|}.
\end{align}

Then the algorithm iterates with $i$ starting at $1$
\begin{align}
 a_{i} &= \elemm{\psi_{i}}{H}{\psi_{i}} \label{eq:ai}\\
 b_{i+1} &= \| H \ket{\psi_{i}} - a_{i} \ket{\psi_{i}} - b_{i} \ket{\psi_{i-1}} 
 \| \label{eq:bi} \\
 \ket{\psi_{i+1}} &= \frac{H \ket{\psi_{i}} - a_{i} \ket{\psi_{i}} - b_{i} 
\ket{\psi_{i-1}}}{b_{i+1}} \label{eq:ci}.
\end{align}

The frequency dependence of the dielectric function is computed in an 
efficient way in terms of the continued fraction
\begin{align}
 \varepsilon_M(\omega) = 1 - \lim_{\boldsymbol{q} \rightarrow 0} 
v(\boldsymbol{q}) \frac{\| P(\boldsymbol{q}) \|^2}{(\omega + i\eta) - a_1 - 
\frac{b_2^2}{(\omega + i \eta) - a_2 - \frac{b_3^2}{\cdots}}}
\label{eq:fraccont}
\end{align}
and the iteration is stopped when $\varepsilon_M(\omega)$ is converged for each 
frequency.

The construction of the Krylov chain Eqs.~(\ref{eq:ai}-\ref{eq:ci}) 
requires 
only the application of the Hamiltonian on different functions or, 
in linear algebra language, simple matrix-vector products.
The computational cost scales as $\mathcal{O}(m N^2)$ with $m$ the number of 
iterations of the Lanczos 
algorithm and $N$ the dimension of the matrix. In our approach, the BSE 
Hamiltonian is expressed in the electron-hole basis thus $N = N_v N_c 
N_k$ where $N_v$ is the number of valence bands, $N_c$ the number of 
conduction bands and $N_k$ the number of points in the BZ.

The number of iterations $m$ needed to converge $\varepsilon_M(\omega)$ is much 
smaller than the size of the Hamiltonian and almost independent of the size of 
the system. As a consequence, Lanczos methods are much more efficient than 
direct diagonalization techniques that scale as $\mathcal{O}(N^3)$.
Unfortunately, unlike diagonalization methods, the Lanczos approach does not 
give direct access to the exciton levels and the corresponding wavefunctions.

The computation of the BSE matrix elements and the storage of the Hamiltonian 
represent the most CPU-intensive and memory demanding parts. For example, a 
converged computation of the 
dielectric function of bulk silicon requires few bands (3-4 valence bands, 4-6 
conduction bands) but a large number of wavevectors in the BZ (from 
14$\times$14$\times$14 to 40$\times$40$\times$40, depending on the 
accuracy required) which means that about $10^3$ to 
$10^4$ 
wavevectors must be sampled. This gives, for sequential computers, from days to 
years of computation, and in terms of memory, matrices of size ranging from 
32928$\times$32928 (ca. 16 GB) to 1536000$\times$1536000 (ca. 34 TB). Such 
huge memory requirements and the corresponding computation time render BSE 
calculations challenging even on modern supercomputers. These issues are even 
more severe when BSE results are used to 
perform resonant Raman scattering calculations that, as illustrated in 
Ref.~\cite{Gillet2013}, require an exceedingly dense BZ sampling.

These two bottlenecks can be reduced by using the technique presented in the 
next section.

\section{Presentation of the interpolation technique \label{sec3}}

The interpolation scheme we propose is based on two meshes 
of wavevectors in the BZ (double-grid technique). Later, we will distinguish 
different
levels of interpolation, all based on this double-grid technique.

To facilitate the discussion, we introduce the following notation.
The coarse mesh contains $\tilde{N}_k$ homogeneous wavevectors, denoted as 
$\tilde{\boldsymbol{k}}$.
The dense mesh contains $N_k = \tilde{N}_k \times N_{div}$ homogeneous 
wavevectors 
obtained by refining the coarse mesh. 
The refining in each direction is done 
by defining equally spaced $n_i$ points in the $i$-th direction.
The wavevectors of the coarse mesh are given by
\begin{align}
\tilde{\boldsymbol{k}}_{(i_1,i_2,i_3)} = i_1 \hat{\boldsymbol{k}}_1 + i_2 
\hat{\boldsymbol{k}}_2 + i_3 \hat{\boldsymbol{k}}_3,
\end{align}
where $i_j$ are integer coordinates and $\hat{\boldsymbol{k}}_j$ are the basis 
vectors of the coarse mesh.

The dense wavevectors have fractional coordinates
\begin{align}
\boldsymbol{k}_{(i_1,i_2,i_3),(j_1,j_2,j_3)} = (i_1 + \frac{j_1}{n_1}) 
\hat{\boldsymbol{k}}_1 + (i_2 + \frac{j_2}{n_2}) \hat{\boldsymbol{k}}_2 + (i_3 
+ 
\frac{j_3}{n_3}) \hat{\boldsymbol{k}}_3 
\end{align}
where $n_1$, $n_2$ and $n_3$ are the number of divisions along the basis 
vectors while $0 \le j_i \le (n_i - 1)$. For the sake of simplicity, we assume 
the same 
number of divisions, $n_1 = n_2 = n_3 = n_{div}$, along the three directions 
and therefore 
$N_{div} = n_{div}^3$.

The neighborhood of a dense point, $N(\boldsymbol{k})$, is defined as the set 
of the eight wavevectors around $\boldsymbol{k}$. The reduced coordinates of 
this set of points are given by 
\begin{align}
 N(\boldsymbol{k}_{(i_1,i_2,i_3),(j_1,j_2,j_3)}) = 
\left\{\tilde{\boldsymbol{k}}_{(i_1,i_2,i_3),(j_1,j_2,j_3)}^{lmn}  \right\} 
\text{ with } 
l,m,n = 0,1 \label{eq:neighborhood}
\end{align}
where $\tilde{\boldsymbol{k}}^{lmn}$ is the $lmn^\text{th}$-neighbor of 
$\boldsymbol{k}$
\begin{align}
 \tilde{\boldsymbol{k}}_{(i_1,i_2,i_3),(j_1,j_2,j_3)}^{lmn} = 
\tilde{\boldsymbol{k}}_{(i_1+l,i_2+m,i_3+n)}.
\end{align}

The dense set of a coarse point, $S(\tilde{\boldsymbol{k}})$, is defined as
\begin{align}
S(\tilde{\boldsymbol{k}}_{(i_1,i_2,i_3)}) = \left\{ 
\boldsymbol{k}_{(i_1,i_2,i_3),(j_1,j_2,j_3)} \right\} \forall (j_1,j_2,j_3).
\label{eq:subspace}
\end{align}

Using these definitions, we can derive the following important relation
\begin{align}
 \sum_{\boldsymbol{k}} \sum_{\tilde{\boldsymbol{k}} \in N(\boldsymbol{k})} 
f(\boldsymbol{k},\tilde{\boldsymbol{k}})&= 
\sum_{\tilde{\boldsymbol{k}}'} \sum_{\boldsymbol{k} \in 
S(\tilde{\boldsymbol{k}'})} \sum_{\tilde{\boldsymbol{k}} \in 
N(\boldsymbol{k})} f(\boldsymbol{k},\tilde{\boldsymbol{k}}) \nonumber \\
&= 
\sum_{\tilde{\boldsymbol{k}}'} \sum_{\boldsymbol{k} \in 
S(\tilde{\boldsymbol{k}'})} \sum_{\tilde{\boldsymbol{k}} \in 
N(\tilde{\boldsymbol{k}'})}f(\boldsymbol{k},\tilde{\boldsymbol{k}}) \nonumber \\
&= 
\sum_{\tilde{\boldsymbol{k}}'} 
\sum_{\tilde{\boldsymbol{k}} \in 
N(\tilde{\boldsymbol{k}'})}
\sum_{\boldsymbol{k} \in 
S(\tilde{\boldsymbol{k}'})} 
f(\boldsymbol{k},\tilde{\boldsymbol{k}})\label{eq:ordersums}
\end{align}
where Eq.~\eqref{eq:neighborhood} has been used. This property will be used 
afterwards to perform the interpolation of data defined on the coarse mesh.

As discussed in the previous section, the Lanczos algorithm is based on 
matrix-vector products. 
In our method, we perform these operations on-the-fly to avoid 
the storage of 
the BSE Hamiltonian in memory. The matrix elements of the kernel are 
interpolated while performing the matrix-vector operations.
% The interpolation can be of different levels, as will be seen later.
As discussed in more detail in the next paragraphs, different levels of 
interpolation can be employed in this part of the algorithm.

Since the periodic parts of the Bloch states at a given wavevector form a 
complete basis set, any wavefunction on the dense mesh can be 
expressed as~\cite{Rohlfing2000}
\begin{align}
\ket{u_{n\boldsymbol{k}}} = \sum_{n'} 
d_{n\boldsymbol{k}}^{n'\tilde{\boldsymbol{k}}} 
\ket{u_{n'\tilde{\boldsymbol{k}}}}
\label{eq:completeness}
\end{align}
where
\begin{align}
 d_{n\boldsymbol{k}}^{n'\tilde{\boldsymbol{k}}} = 
\prodbk{u_{n'\tilde{\boldsymbol{k}}}}{u_{n\boldsymbol{k}}}.
\end{align}

In the transition basis set, the electron-hole wavefunction 
$\Psi(\boldsymbol{r},\boldsymbol{r}')$ is given by a linear combination of 
products of single-particle orbitals according to
\begin{align}
 \Psi(\boldsymbol{r},\boldsymbol{r}') = \sum_{vc\boldsymbol{k}} 
A_{vc\boldsymbol{k}} \phi_{vc\boldsymbol{k}}(\boldsymbol{r},\boldsymbol{r}')
\end{align}
where 
\begin{align}
 \phi_{vc\boldsymbol{k}}(\boldsymbol{r},\boldsymbol{r}') = e^{-i 
\boldsymbol{k}.\boldsymbol{r}} u^*_{v\boldsymbol{k}}(\boldsymbol{r}) e^{i 
\boldsymbol{k}.\boldsymbol{r}'} u_{c\boldsymbol{k}}(\boldsymbol{r}') = e^{i 
\boldsymbol{k}.(\boldsymbol{r}' - \boldsymbol{r})} 
U_{vc\boldsymbol{k}}(\boldsymbol{r},\boldsymbol{r}').
\end{align}

One basis function on the 
dense mesh can then be 
expanded in term of the wavefunctions located on a coarse point 
by means of
\begin{align}
\ket{U_{vc\boldsymbol{k}}} = \sum_{n_1 n_2} 
(d_{v\boldsymbol{k}}^{n_1\tilde{\boldsymbol{k}}})^*
d_{c\boldsymbol{k}}^{n_2\tilde{\boldsymbol{k}}} \ket{U_{n_1 n_2 
\tilde{\boldsymbol{k}}}}. \label{eq:expansion}
\end{align}

The method developed by Rohlfing and Louie in Ref.~\cite{Rohlfing2000} 
uses a single reference point $\tilde{\boldsymbol{k}}$ to expand the 
kernel according to
\begin{align}
 K^i_{vc\boldsymbol{k},v'c'\boldsymbol{k}'} =  \sum_{n_1 n_2} 
d_{v\boldsymbol{k}}^{n_1\tilde{\boldsymbol{k}}} 
(d_{c\boldsymbol{k}}^{n_2\tilde{\boldsymbol{k}}})^*
  \sum_{n_3 n_4} 
 (d_{v'\boldsymbol{k}'}^{n_3\tilde{\boldsymbol{k}'}})^*
 d_{c'\boldsymbol{k}'}^{n_4\tilde{\boldsymbol{k}'}}  
 K_{n_1 n_2 \tilde{\boldsymbol{k}}, n_3 n_4 
\tilde{\boldsymbol{k}'}} \label{eqRL}
\end{align}
and we generalize their approach by including eight coarse 
points in the 
expansion of the wavefunctions
\begin{align}
\ket{U_{vc\boldsymbol{k}}} = \sum_{\tilde{\boldsymbol{k}} \in 
N(\boldsymbol{k})} f(\boldsymbol{k},\tilde{\boldsymbol{k}}) \sum_{n_1 
n_2} (d_{v\boldsymbol{k}}^{n_1\tilde{\boldsymbol{k}}})^* 
d_{c\boldsymbol{k}}^{n_2\tilde{\boldsymbol{k}}} 
\ket{U_{n_1 n_2 \tilde{\boldsymbol{k}}}}, \label{eq:interpwfn}
\end{align}
where $f(\boldsymbol{k},\tilde{\boldsymbol{k}})$ are interpolation prefactors. 
The RL interpolation scheme is a special case of Eq.~\eqref{eq:interpwfn} in 
which $f(\boldsymbol{k},\tilde{\boldsymbol{k}}) = 
1$ for a chosen neighbor and 0 for all the other ones. 

In order to accelerate the convergence of the expansion, we perform a trilinear 
interpolation of the coefficients. In this case, the prefactors are given by 
\begin{align}
 f(\boldsymbol{k},\tilde{\boldsymbol{k}}) &= 0 &\text{ if } 
\tilde{\boldsymbol{k}} \not\in N(\boldsymbol{k}) \nonumber \\
 f(\boldsymbol{k},\tilde{\boldsymbol{k}}^{lmn}) &= 
f^{lmn}_{\boldsymbol{k}_{(i_1,i_2,i_3),(j_1,j_2,j_3)}} = f^l_{j_1} 
f^m_{j_2} f^n_{j_3}
& 
\end{align}
with
\begin{equation}
 f^{l}_j = \left\{ \begin{split}
                 1-\frac{j}{n_{div}} ~ \text{if} ~ l = 0 \\
                 \frac{j}{n_{div}} ~ \text{if} ~ l = 1.
                \end{split} \right. 
\end{equation}

Using Eq.~\eqref{eq:interpwfn}, one obtains the following expression for the 
interpolated matrix elements
\begin{align}
  K^i_{vc\boldsymbol{k},v'c'\boldsymbol{k}'} = 
\sum_{\tilde{\boldsymbol{k}} \in N(\boldsymbol{k})}
f(\boldsymbol{k},\tilde{\boldsymbol{k}})
  \sum_{n_1 n_2} d_{v \boldsymbol{k}}^{n_1,\tilde{\boldsymbol{k}}} 
(d_{c\boldsymbol{k}}^{n_2\tilde{\boldsymbol{k}}})^*
   \sum_{\tilde{\boldsymbol{k}}' \in N(\boldsymbol{k}')}
 f(\boldsymbol{k}',\tilde{\boldsymbol{k}}') \sum_{n_3 n_4} 
 (d_{v'\boldsymbol{k}'}^{n_3\tilde{\boldsymbol{k}}'})^* 
 d_{c'\boldsymbol{k}'}^{n_4\tilde{\boldsymbol{k}}'} 
 K_{n_1 n_2 \tilde{\boldsymbol{k}}, n_3 n_4 
\tilde{\boldsymbol{k}'}} \label{interp1}.
\end{align}

By using the overlaps of the periodic parts of the 
wavefunctions, we are thus able to include correctly the phases of the 
wavefunctions 
and these phases will cancel out with the oscillator matrix elements 
$P(\boldsymbol{q})$ computed with the wavefunctions on the dense mesh.

It should be stressed, however, that the matrix 
elements of the Coulomb interaction 
diverge when $\boldsymbol{q} = \boldsymbol{k} - \boldsymbol{k}' \rightarrow
0$. Following Ref.~\cite{Rohlfing2000}, we rewrite the matrix elements as
\begin{align}
 K_{v c \boldsymbol{k}, v' c' 
 \boldsymbol{k}'} = 
\frac{a_{vc\boldsymbol{k},v'c'\boldsymbol{k}'}}{\boldsymbol{q}^2} + 
\frac{b_{vc\boldsymbol{k},v'c'\boldsymbol{k}'}}{\boldsymbol{q}} +
c_{vc\boldsymbol{k},v'c'\boldsymbol{k}'} \label{diverg}
\end{align}
and we note that an accurate interpolation technique should try to reproduce 
the divergent behavior as much as possible.

The different schemes we have implemented to treat the divergence are 
discussed in more detail in the next section.

\section{Combining Lanczos algorithm with interpolation \label{sec4}}

As previously discussed, the dimension of the matrix on the coarse mesh is 
$N_{coarse} = N_c N_v \tilde{N}_k$, while the dimension of the matrix on the 
dense mesh is $N_{dense} = N_c N_v N_k = N_c N_v \tilde{N}_k N_{div}$.
The calculation of the matrix elements of the Hamiltonian as well as the 
Lanczos algorithm scale as $\mathcal{O}(N^2)$. The numerical complexity of the 
standard 
BSE solution on the coarse mesh is thus 
\begin{align}
 \mathcal{O}(N^2_c N^2_v \tilde{N}^2_k)
\end{align}
while the complete solution on the dense mesh scales as
\begin{align}
 \mathcal{O}(N^2_c N^2_v \tilde{N}^2_k  N^2_{div}).
\end{align}

To fix the ideas, supposing a halving of the coarse mesh for the three 
directions giving the dense mesh, $N_{div}=8$ and $N_{div}^2=64$, which points 
out the significant burden of using the dense meshes. If $\tilde{N}_k$ is kept 
constant, and $N_{div}$ is increased, the use of the dense mesh is even more 
unfavorable.

The most memory-demanding part is the storage of the Hamiltonian which scales 
quadratically with the size of the Hamiltonian.  The interpolation technique 
given in Eq.~\eqref{interp1} can be 
implemented 
in two different ways. The interpolated matrix elements 
can be stored in memory and then used as a standard matrix for the Lanczos 
technique. This is the approach followed by Rohlfing in \cite{Rohlfing2000}. It 
is 
worth noting, however, that although the RL method 
allows one to avoid the explicit computation of the matrix elements on the dense 
mesh, the numerical complexity and the memory requirements of the approach are 
still the ones of a standard BSE.

Alternatively, one can reformulate the equations so that the interpolation is 
done on-the-fly without allocating extra memory for the dense Hamiltonian. This 
is the central result of this paper. As the Lanczos technique 
requires only matrix-vector multiplications, the 
full-matrix vector multiplication with the Hamiltonian can be written as
\begin{align}
 \phi^{(n+1)}_{vc\boldsymbol{k}} =& \sum_{v'c'\boldsymbol{k}'} 
H^i_{vc\boldsymbol{k},v'c'\boldsymbol{k}'} \phi^{(n)}_{v'c'\boldsymbol{k}'} \\
=& \left( \varepsilon_{c\boldsymbol{k}} 
- \varepsilon_{v\boldsymbol{k}} \right) \phi^{(n)}_{vc\boldsymbol{k}} + 
\sum_{v'c'\boldsymbol{k}'} 
K^i_{vc\boldsymbol{k},v'c'\boldsymbol{k}'} \phi^{(n)}_{v'c'\boldsymbol{k}'}
\end{align}
that can be computed with $\mathcal{O}(N_c N_v \tilde{N}_k 
N_{div})$ scaling. The matrix-vector product with the kernel can be 
rewritten as
\begin{align}
 s_{vc\boldsymbol{k}} =& \sum_{v'c'\boldsymbol{k}'} 
K^i_{vc\boldsymbol{k},v'c'\boldsymbol{k}'} p_{v'c'\boldsymbol{k}'} \\
 =& \sum_{\tilde{\boldsymbol{k}} \in N(\boldsymbol{k})}
f(\boldsymbol{k},\tilde{\boldsymbol{k}})
  \sum_{n_1 n_2} d_{v \boldsymbol{k}}^{n_1,\tilde{\boldsymbol{k}}} 
(d_{c\boldsymbol{k}}^{n_2\tilde{\boldsymbol{k}}})^* \nonumber \\
 & \sum_{\tilde{\boldsymbol{k}}''} \sum_{\tilde{\boldsymbol{k}}' \in 
N(\tilde{\boldsymbol{k}}'')} \sum_{n_3 n_4} 
 K_{n_1 n_2 \tilde{\boldsymbol{k}}, n_3 n_4 
\tilde{\boldsymbol{k}'}}  \nonumber \\
& \sum_{\boldsymbol{k}' \in 
S(\tilde{\boldsymbol{k}}'')}
 f(\boldsymbol{k}',\tilde{\boldsymbol{k}}') \sum_{v'c'}  
 (d_{v'\boldsymbol{k}'}^{n_3\tilde{\boldsymbol{k}}'})^* 
 d_{c'\boldsymbol{k}'}^{n_4\tilde{\boldsymbol{k}}'} p_{v'c'\boldsymbol{k}'}  
\label{interpmatmul}.
\end{align}
and can be computed in three steps using
\begin{align}
 q^{uvw}_{n_3 n_4 \tilde{\boldsymbol{k}''}} =&
 \sum_{\boldsymbol{k}' \in 
S(\tilde{\boldsymbol{k}}'')}
 f(\boldsymbol{k}',\tilde{\boldsymbol{k}}') \sum_{v'c'}  
 (d_{v'\boldsymbol{k}'}^{n_3\tilde{\boldsymbol{k}}'})^* 
 d_{c'\boldsymbol{k}'}^{n_4\tilde{\boldsymbol{k}}'} p_{v'c'\boldsymbol{k}'} 
\text{ with } \tilde{\boldsymbol{k}}' = {\tilde{\boldsymbol{k}''}}^{uvw}
\label{matmul1} \\
 r_{n_1 n_2 \tilde{\boldsymbol{k}}} =& \sum_{\tilde{\boldsymbol{k}}''} 
\sum_{uvw} \sum_{n_3 n_4} 
 K_{n_1 n_2 \tilde{\boldsymbol{k}}, n_3 n_4 
({\tilde{\boldsymbol{k}''}}^{uvw})} q^{uvw}_{n_3 n_4 \tilde{\boldsymbol{k}''}} 
\label{matmul2} \\
 s_{vc\boldsymbol{k}} =& \sum_{\tilde{\boldsymbol{k}} \in N(\boldsymbol{k})}
f(\boldsymbol{k},\tilde{\boldsymbol{k}})
  \sum_{n_1 n_2} d_{v \boldsymbol{k}}^{n_1,\tilde{\boldsymbol{k}}} 
(d_{c\boldsymbol{k}}^{n_2\tilde{\boldsymbol{k}}})^* r_{n_1 n_2 
\tilde{\boldsymbol{k}}} 
\label{matmul3}.
\end{align}

\begin{figure}
 \centering
 \def\svgwidth{7cm}
 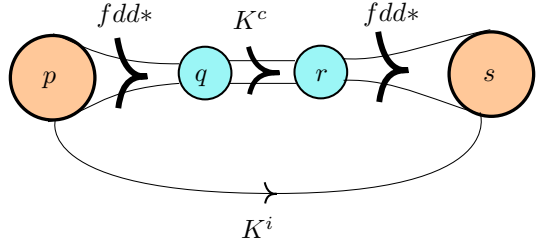
 \caption{Color online. Schema illustrating the interpolated matrix-vector 
product. Small circles represent coarse mesh data while big circles correspond 
to dense
mesh data. $fdd*$ refers to the $f(\boldsymbol{k},\tilde{\boldsymbol{k}})
  d_{v \boldsymbol{k}}^{n_1,\tilde{\boldsymbol{k}}} 
(d_{c\boldsymbol{k}}^{n_2\tilde{\boldsymbol{k}}})^*$ prefactors of 
Eq.~\eqref{matmul1} and Eq.~\eqref{matmul3}. $K_c$ is the application of the 
coarse kernel on the coarse vector $q$ that gives $r$ in Eq.~\eqref{matmul2}.
\label{fig:schema-interp}}
\end{figure}

A schematic representation of the algorithm is given in 
Fig.~\ref{fig:schema-interp}.
The application of the interpolated Hamiltonian is equivalent 
to averaging dense vector ($p$) on a coarse vector ($q$), then applying the 
coarse 
Hamiltonian ($r$) and finally rebuilding the full vector information on the 
dense 
mesh ($s$).

The numerical complexity of Eq.~\eqref{matmul1}, \eqref{matmul2} and 
\eqref{matmul3} is $\mathcal{O}(N^2_c N^2_v \tilde{N}_k N_{div})$, 
$\mathcal{O}(N^2_c N^2_v \tilde{N}^2_k)$, and $\mathcal{O}(N^2_c N^2_v 
\tilde{N}_k N_{div})$ respectively instead of the $\mathcal{O}(N^2_c N^2_v 
\tilde{N}^2_k  N^2_{div})$ scaling of a BSE run done on the same dense mesh 
without interpolation. This first approach is called ``Method 1'' (M1) in 
the rest of this work.
 
As stated at the end of Section~\ref{sec3}, a better treatment of the 
divergence is expected to improve the accuracy of the 
interpolation. Considering Eq.~\eqref{diverg}, one can interpolate 
the coefficients and then divide the interpolated quantities by the 
$\boldsymbol{q}$ computed on the dense mesh.
The drawback of this approach, however, is that it requires the computation of 
the whole matrix since the fast algorithm developed in 
Eq.~\eqref{interpmatmul} is not applicable thus resulting in $\mathcal{O}(N_c^2 
N_v^2 \tilde{N}_k^2 N_{div}^2)$ scaling.
This approach is called ``Method 2'' (M2) in the rest of this work. 

\begin{figure}
 \centering
 \def\svgwidth{4cm}
 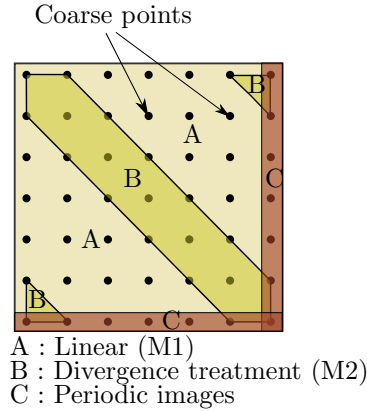
 \caption{Regions of the Hamiltonian where the interpolation is applied in 
``Method 3'', for a width $w = 1.0$.}
 \label{figdiv}
\end{figure}

The last method (``Method 3'', M3) has been developed as a compromise between 
accuracy and 
numerical complexity.
In this case, the divergent behavior is reproduced only in a small region along 
the diagonal with a width that can 
be adjusted by the user (see Fig.~\ref{figdiv}). This approximation allows us 
to employ the fast interpolation of Eq.~\eqref{interpmatmul} for the full 
matrix. The computational cost needed to treat the divergence is negligible 
provided that the width is small with respect to the number of points on the 
coarse mesh. Under this assumption, the overall 
complexity of M3 is $\mathcal{O}(N_c^2 N_v^2 \tilde{N}_k N_{div}^2)$. 

We define the width $w$ relatively to the smallest distance between two 
points in the coarse grid $d$. Then, all $(k,k')$ pairs in the 
dense mesh so that $||k-k'|| > w \times d$ are treated with Method 1 and for 
the other pairs, the coefficients of Eq.~\eqref{diverg} are interpolated and 
used together with the dense $\boldsymbol{q}$ to treat the divergence.

\section{Comparison of the interpolation schemes \label{sec5}}

In this section, the different interpolation schemes are tested and compared in 
detail. Our method has been implemented in the open source ABINIT 
code~\cite{Gonze2005,Gonze2009} and will be made available in the forthcoming 
release. First, three 
different prototype systems, silicon, gallium arsenide and lithium fluoride, 
are studied and the accuracy of the three schemes discussed in the previous 
section is studied in detail. 
Then, a non-physical test case with very low convergence parameters is used to 
analyze how the computational cost scales with the total number of points 
employed
to sample the BZ.

\subsection{Accuracy on test cases}

Silicon and gallium arsenide have relatively high dielectric 
constant (10.9 for GaAs and 12 for Si~\cite{Yu2010}) and therefore small 
binding energies (4-5 meV for GaAs~\cite{Rohlfing1998,Yu2010} and 15 meV for 
Si~\cite{Green2013}) and Mott-Wannier-like excitons. LiF, on the other hand, 
has a relatively small dielectric constant of 1.9~\cite{Rohlfing1998}, yielding 
a weak screened interaction and therefore strong excitonic 
effects 
(binding energy on the order of 
3~eV~\cite{Benedict1998a,Rohlfing1998,Arnaud2001,Marini2003}), and Frenkel-like 
excitons. We decided to use these prototype semi-conductors because, as 
discussed in~\cite{Sander2015}, their BSE matrices have very 
different behavior in k-space and it is important to understand how our 
interpolation scheme performs in two different scenarios.

Si and GaAs have been simulated using cut-off energies of 16~Ha 
for 
the wavefunctions and 4~Ha for the dielectric matrix, while for LiF, a cut-off 
energy of 50~Ha has been used for the wavefunctions and 4 Ha for the 
dielectric matrix. Three valence bands and four 
conduction bands were included in the electron-hole basis set. Lanczos chain 
iterations were stopped when the full 
dielectric spectrum reached a maximum relative difference of $1\%$ both on the 
real 
part and the imaginary part.
The model dielectric function of Ref.~\cite{Cappellini1993} has been used to 
avoid the computation 
of the inverse dielectric matrix, with the parameter $\epsilon^\infty$ set 
to 12, 10 and 2 for Si, GaAs and LiF 
respectively. A scissors shift is applied on top of the LDA Kohn-Sham 
eigenvalues to mimic the effect of the GW approximation (0.8 eV for Si and GaAs, 
 5.7 eV for LiF). The broadening factor (see Eq.~\eqref{epsmacro}) is $\eta = 
0.1$~eV. 

\begin{figure}
 \subfloat[$4 \times 4 \times 4$ with 1 neighbor]{ 
 \def\svgwidth{7cm}
 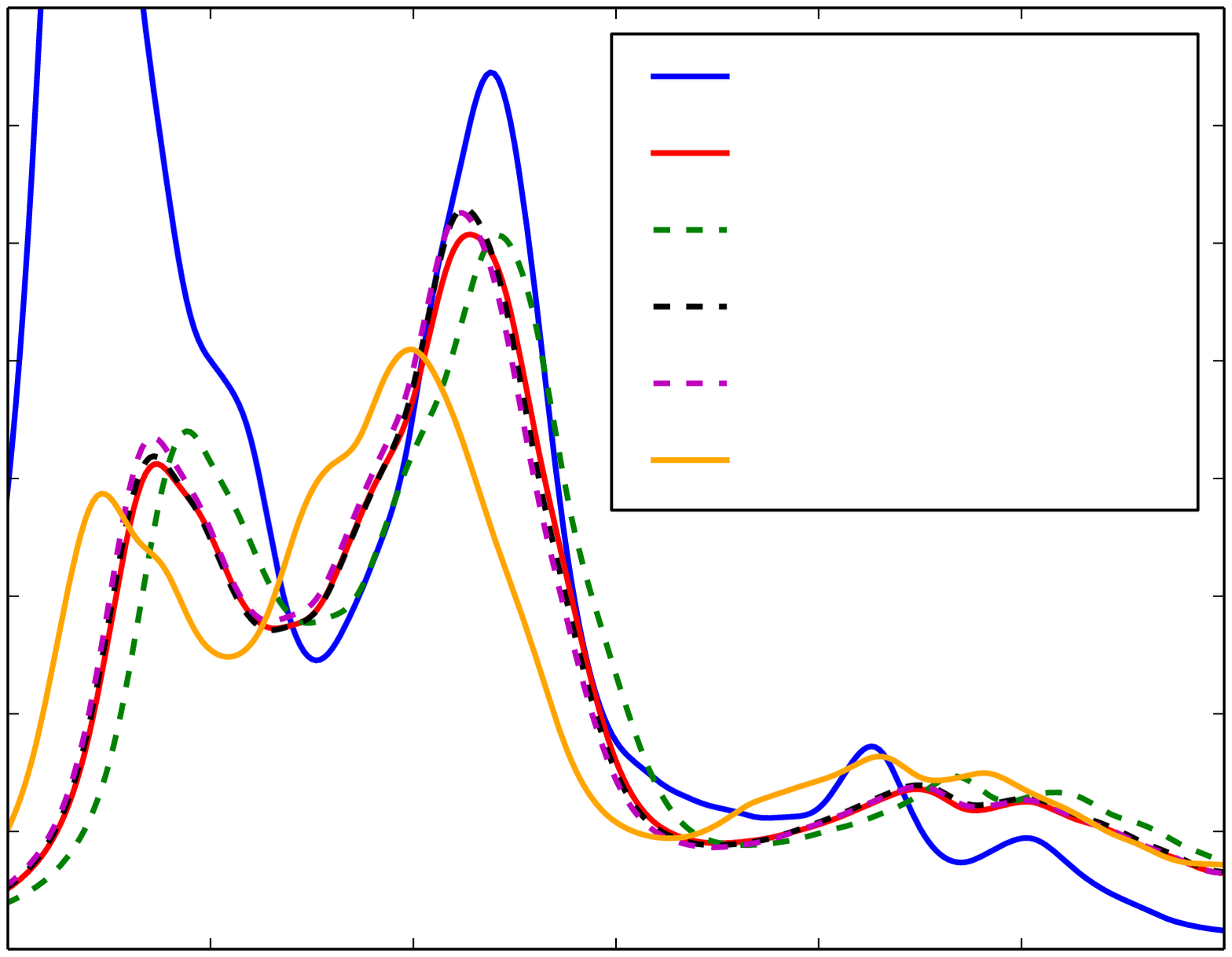}
 \subfloat[$4 \times 4 \times 4$ with 8 neighbors]{
 \def\svgwidth{7cm}
 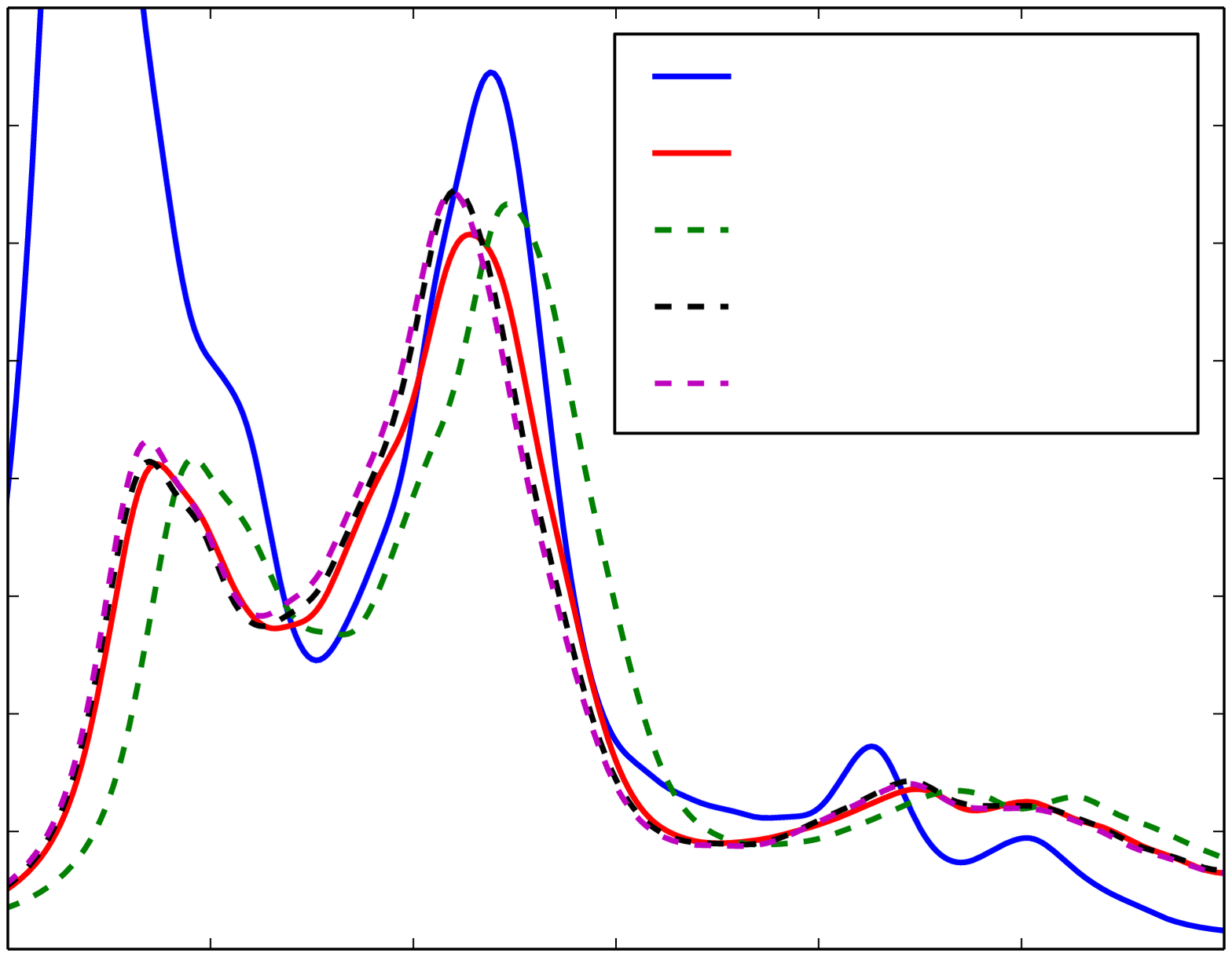} \\
 \subfloat[$8 \times 8 \times 8$ with 1 neighbor]{
 \def\svgwidth{7cm}
 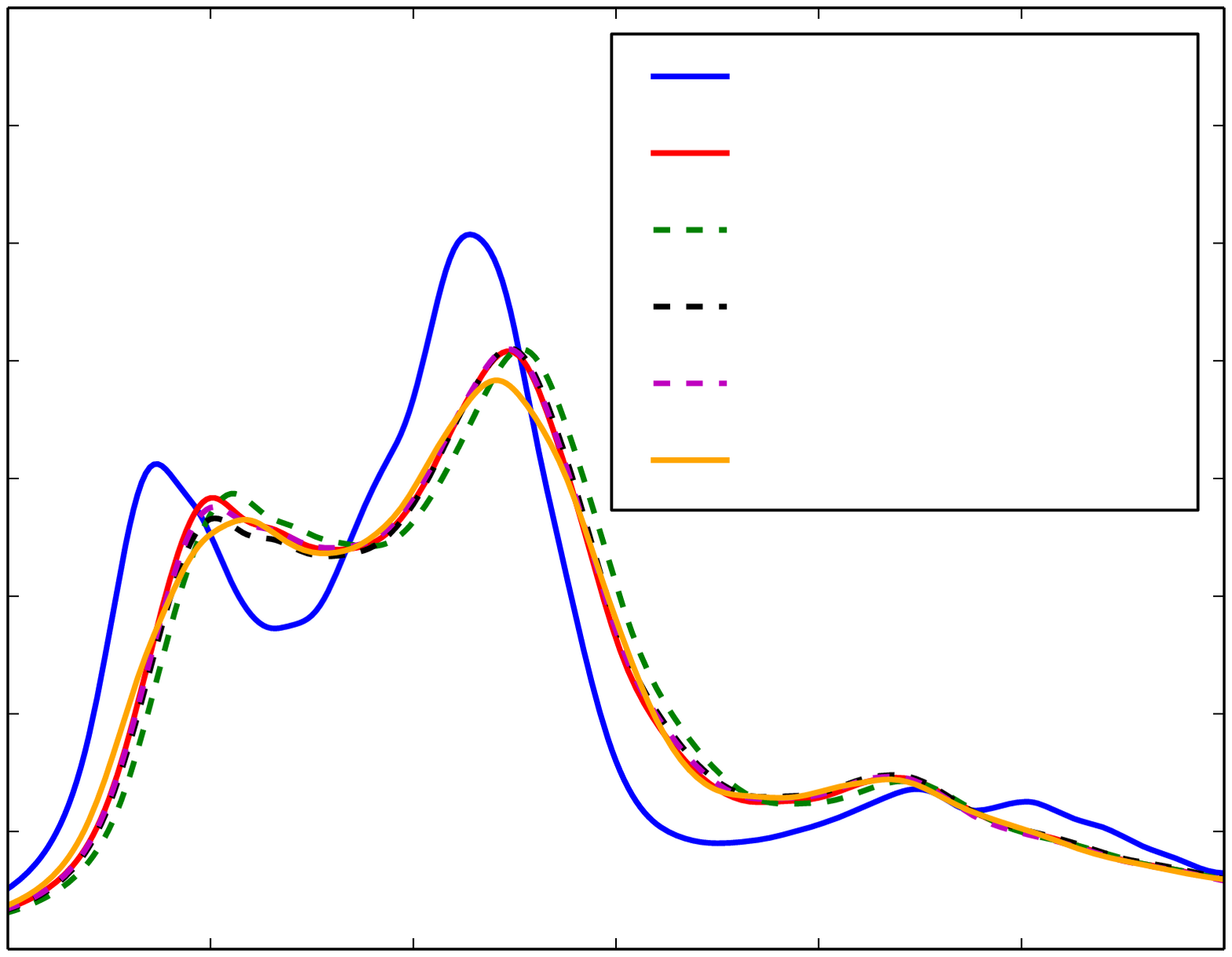}
 \subfloat[$8 \times 8 \times 8$ with 8 neighbors]{
 \def\svgwidth{7cm}
 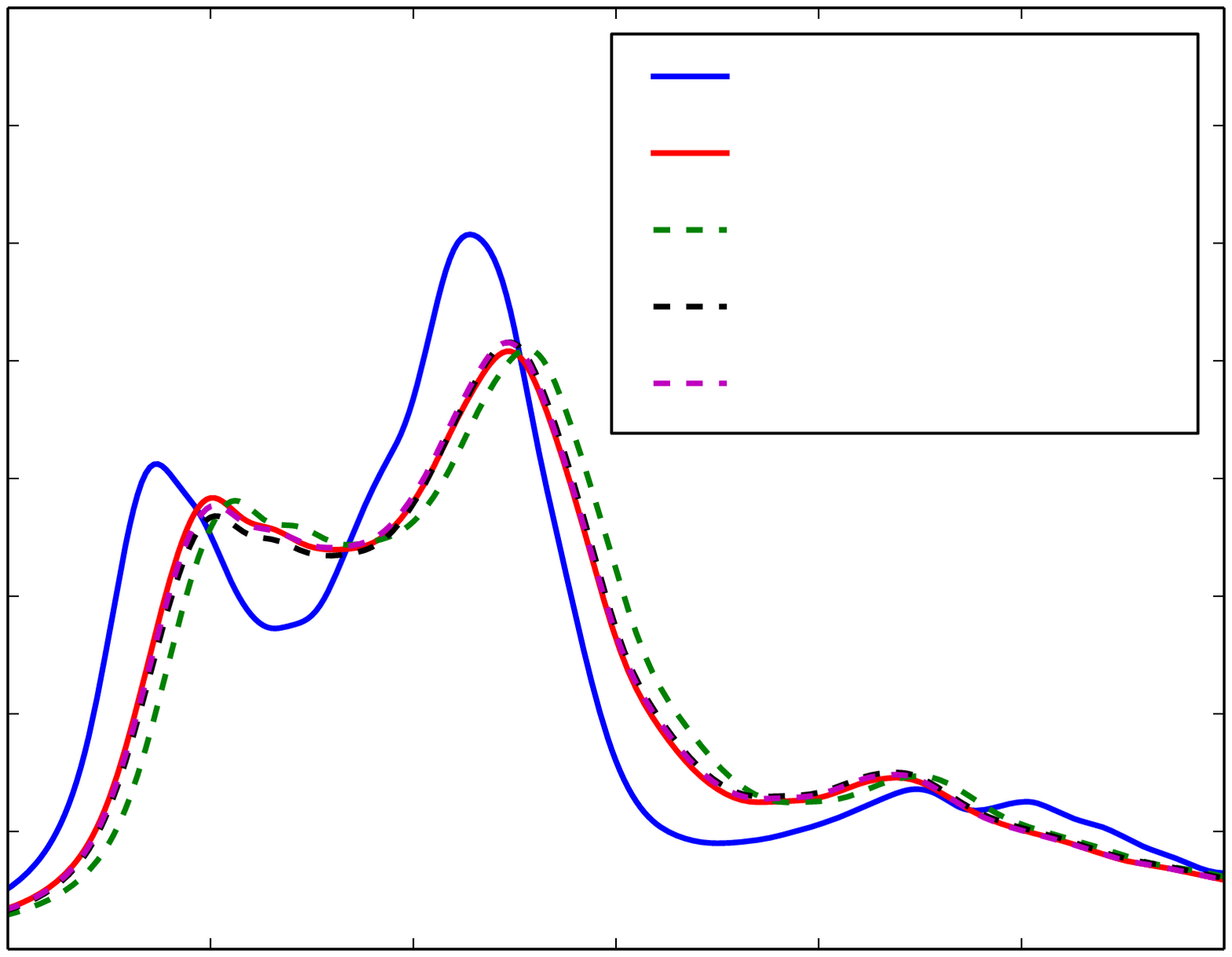}
 \caption{(Color online). Comparison between the absorption spectra of silicon 
obtained with the
standard (non-interpolated) BSE solution and with the six levels 
of interpolation developed in this work. (1NB) refers to the 1-neighbor Rohlfing 
and Louie technique, whose results 
are presented in the left column. (8NB) refers 
to the multilinear technique with 8 neighbors presented in this article, whose 
results are presented in the right column.  
Ref.~\cite{Gillet2013} refers to the multiple-shift technique. The results 
presented in the upper row correspond to the interpolation from the 4x4x4 
grid to the 8x8x8 grid,
while the lower row corresponds to the interpolation from the 8x8x8 grid to the 
16x16x16 grid.}
 \label{fig:interpsi}
\end{figure}

\begin{figure}
 \subfloat[$4 \times 4 \times 4$ with 1 neighbor]{ 
 \def\svgwidth{7cm}
 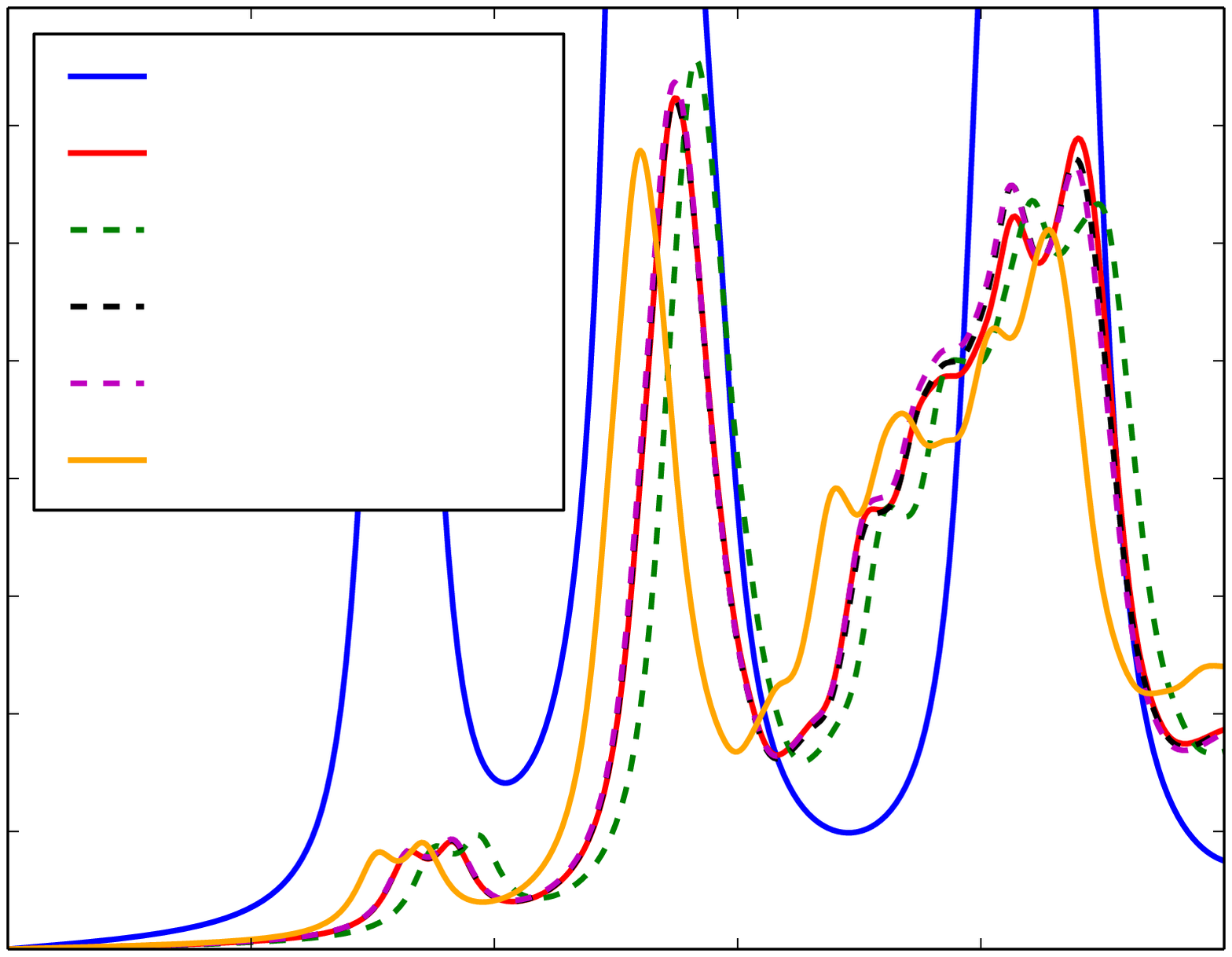}
 \subfloat[$4 \times 4 \times 4$ with 8 neighbors]{
 \def\svgwidth{7cm}
 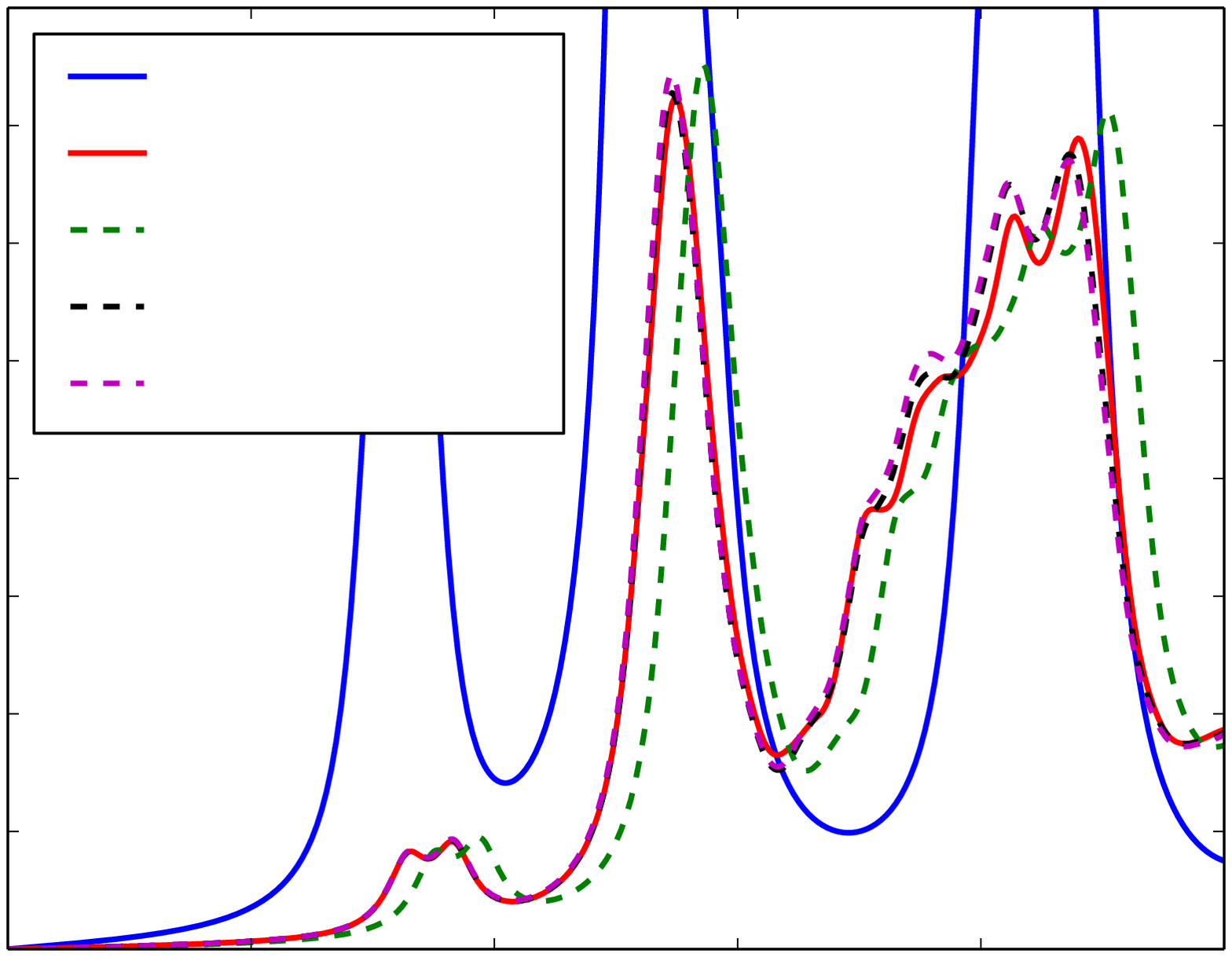} \\
 \subfloat[$8 \times 8 \times 8$ with 1 neighbor]{
 \def\svgwidth{7cm}
 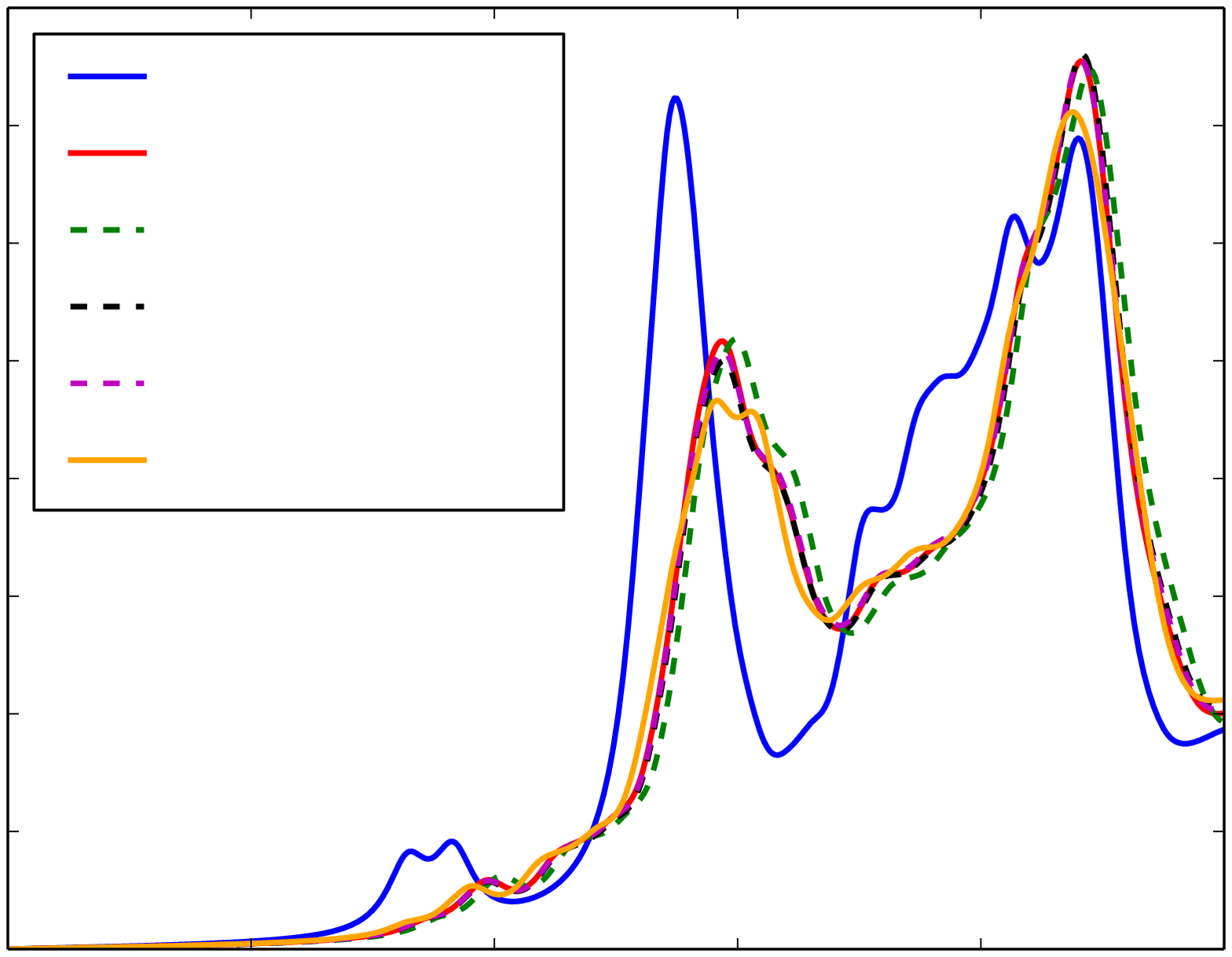}
 \subfloat[$8 \times 8 \times 8$ with 8 neighbors]{
 \def\svgwidth{7cm}
 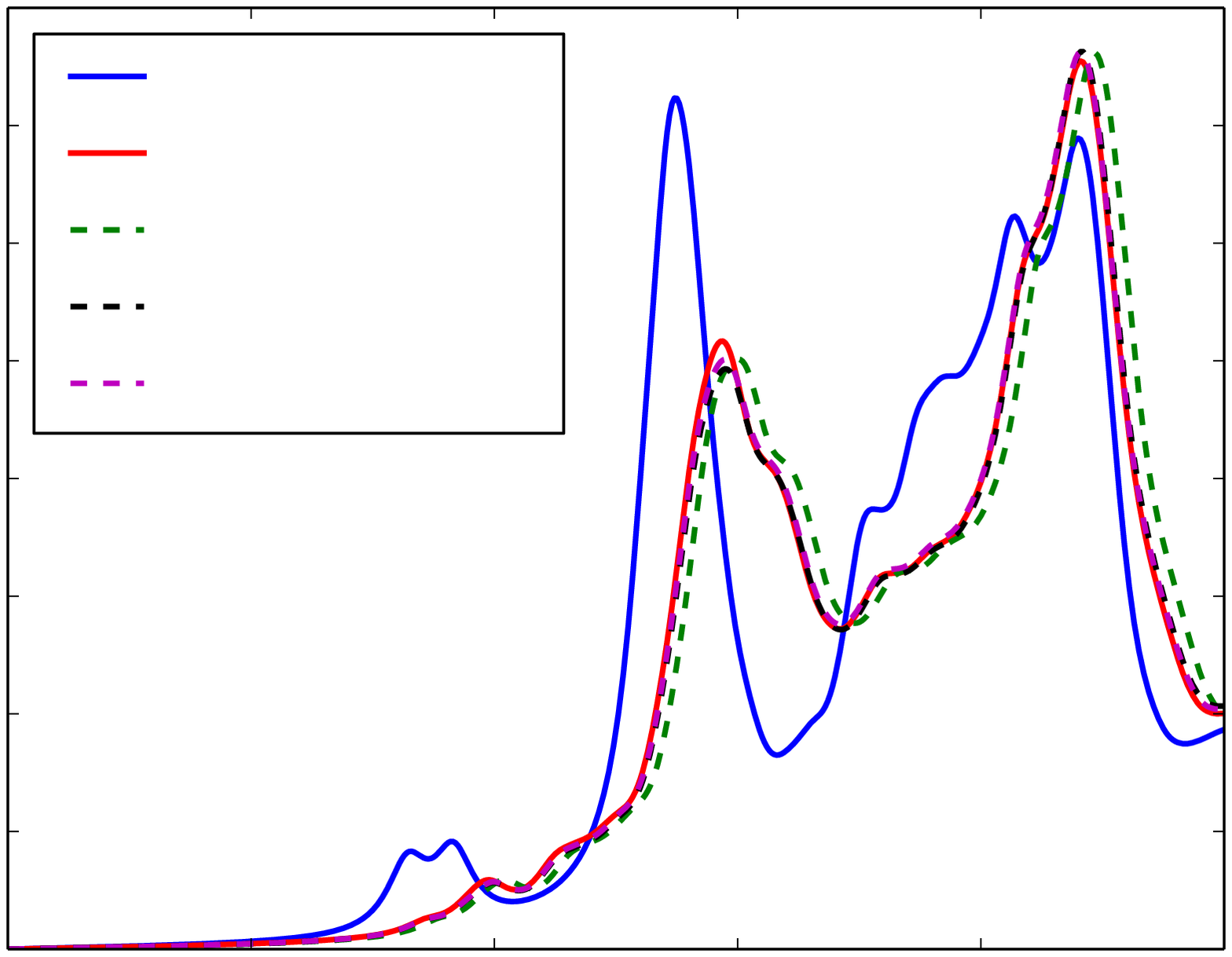}
  \caption{(Color online). Comparison between the absorption spectra of gallium 
arsenide 
obtained with the 
standard (non-interpolated) BSE solution and with the six levels of 
interpolation developed in this 
work. (1NB) refers to the 1-neighbor Rohlfing and Louie technique, whose 
results 
are presented in the left column. (8NB) refers 
to the multilinear technique with 8 neighbors presented in this article, whose 
results are presented in the right column.  
Ref.~\cite{Gillet2013} refers to the multiple-shift technique. The results 
presented in the upper row correspond to the interpolation from the 4x4x4 
grid to the 8x8x8 grid,
while the lower row corresponds to the interpolation from the 8x8x8 grid to the 
16x16x16 grid.}
 \label{fig:interpgaas}
\end{figure}

\begin{figure}
 \subfloat[$4 \times 4 \times 4$ with 1 neighbor]{
 \def\svgwidth{7cm}
 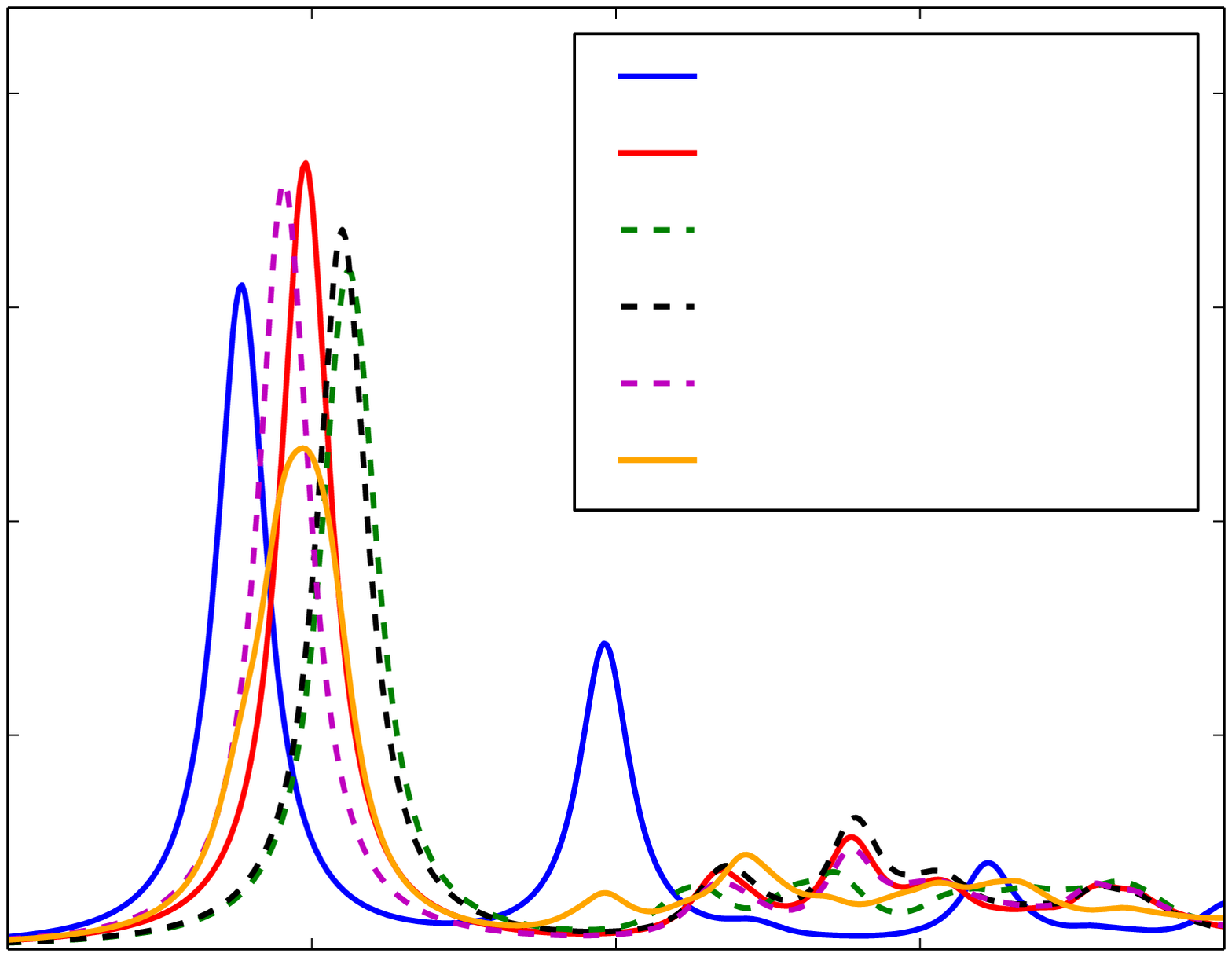}
 \subfloat[$4 \times 4 \times 4$ with 8 neighbors]{
 \def\svgwidth{7cm}
 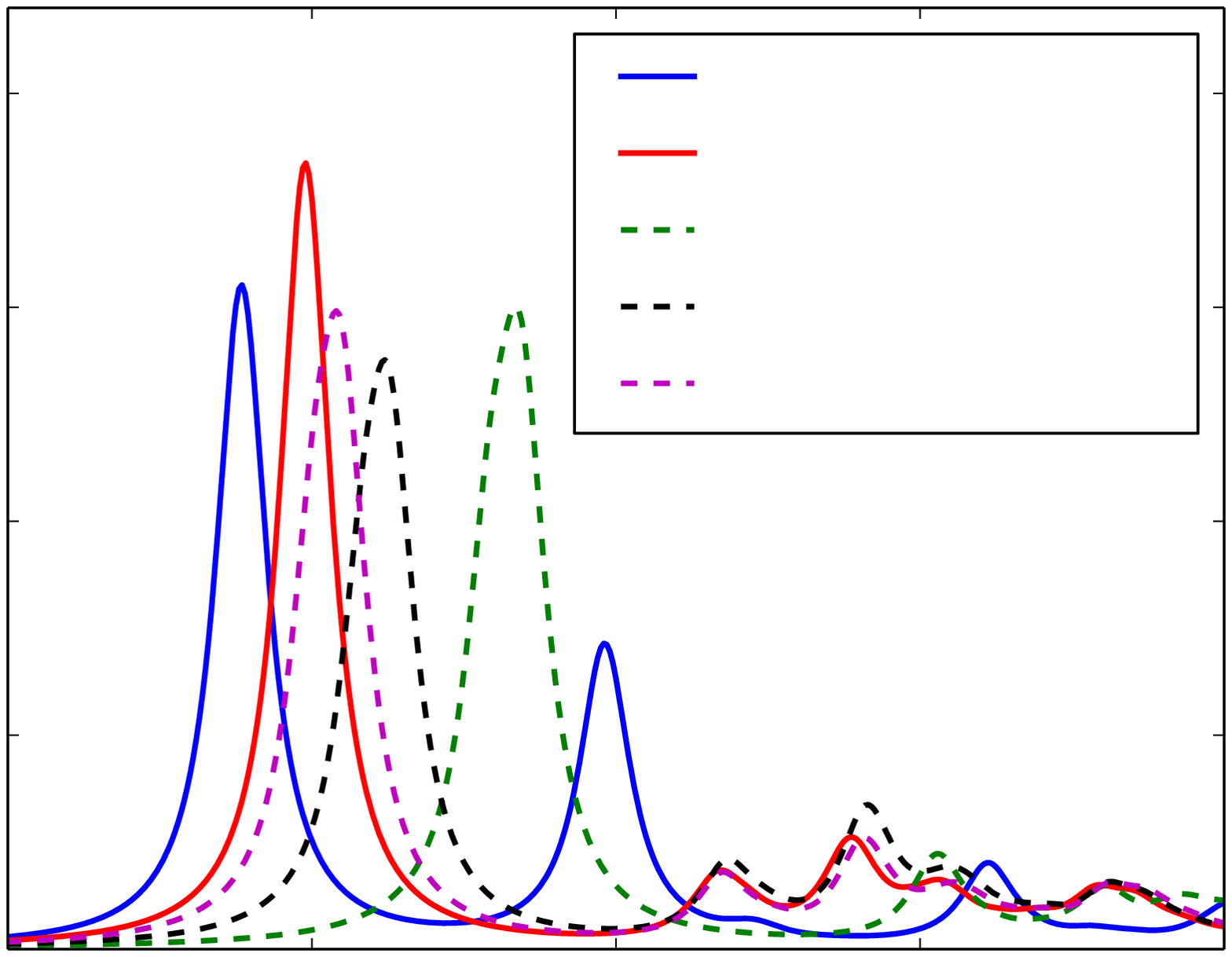}
 \\
 \subfloat[$8 \times 8 \times 8$ with 1 neighbor]{
 \def\svgwidth{7cm}
 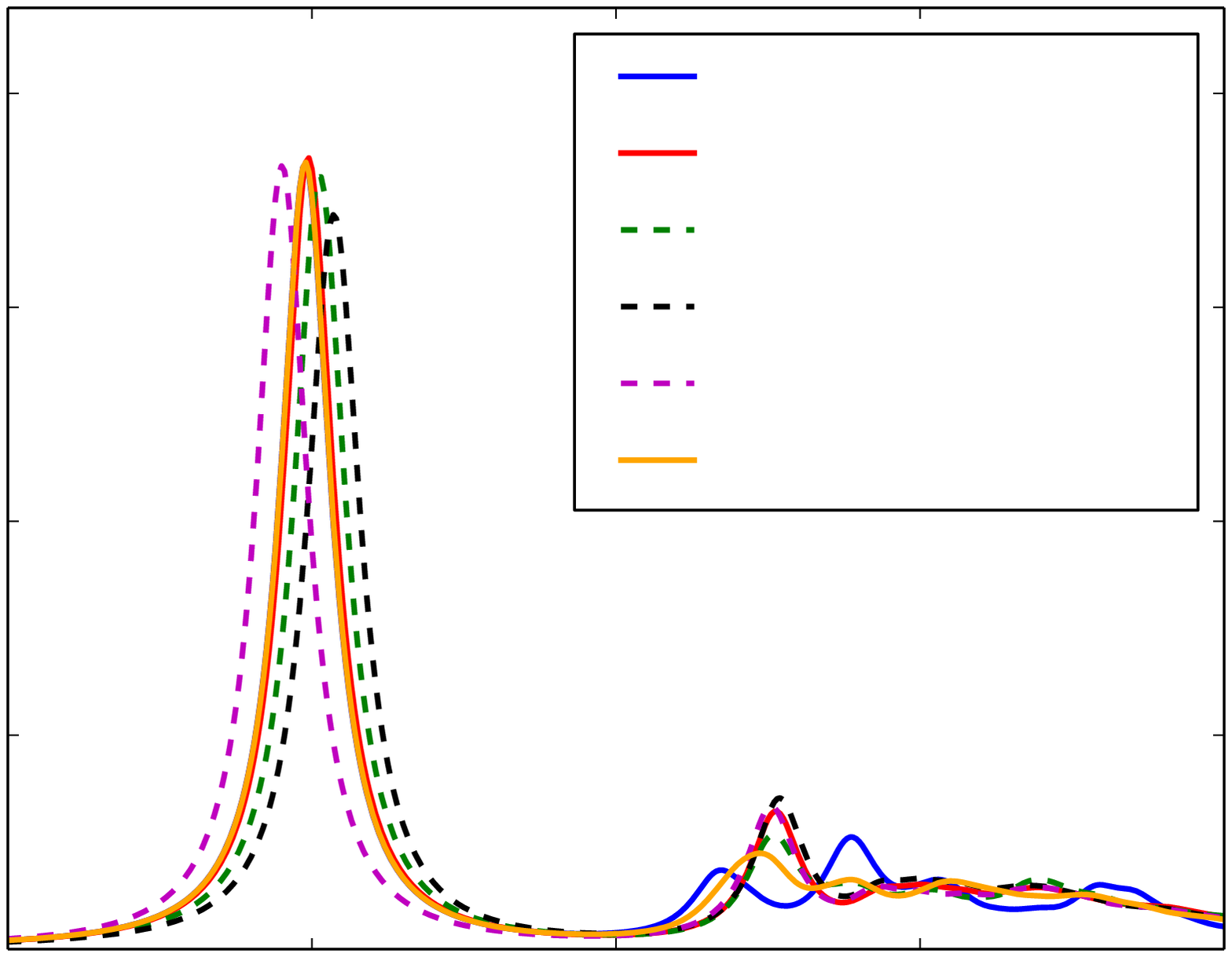}
 \subfloat[$8 \times 8 \times 8$ with 8 neighbors]{
 \def\svgwidth{7cm}
 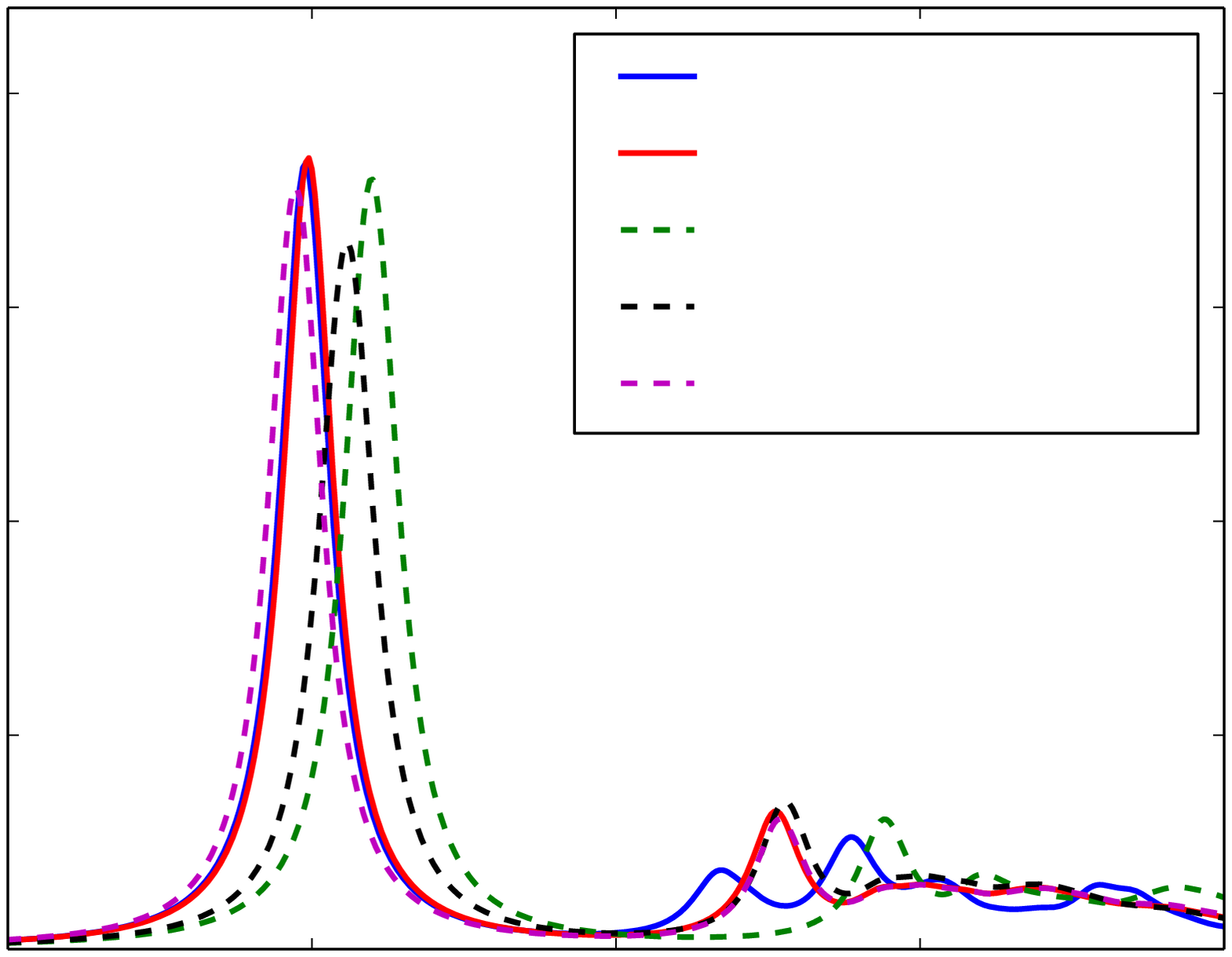}
 \caption{(Color online). Comparison between the absorption spectra of lithium 
fluoride
obtained with the 
standard (non-interpolated) BSE solution and with the six levels of 
interpolation developed in this 
work. (1NB) refers to the 1-neighbor Rohlfing and Louie technique, whose 
results 
are presented in the left column. (8NB) refers 
to the multilinear technique with 8 neighbors presented in this article, whose 
results are presented in the right column.  
Ref.~\cite{Gillet2013} refers to the multiple-shift technique. The results 
presented in the upper row correspond to the interpolation from the 4x4x4 
grid to the 8x8x8 grid,
while the lower row corresponds to the interpolation from the 8x8x8 grid to the 
16x16x16 grid.}
 \label{fig:interplif}
\end{figure}

Results with two coarse grids of $4\times4\times4$ and $8\times8\times8$, 
interpolated to  $8\times8\times8$ and $16\times16\times16$, respectively, are 
presented 
in Fig. \ref{fig:interpsi}, \ref{fig:interpgaas} and \ref{fig:interplif} for 
silicon, gallium arsenide and lithium fluoride respectively. 
The three main peak positions and maximum amplitudes extracted from these 
results are presented in Table.~\ref{tab:interpsi}, \ref{tab:interpgaas} and 
\ref{tab:interplif}.
All the 
calculations are done with BZ meshes shifted along the $(0.011, 0.021, 0.031)$ 
direction in order to improve the accuracy of the sampling.

\begin{table}
 \centering
 \begin{tabular}{|c|cc|cc|cc|}
  \hline
  \multicolumn{1}{|c|}{\bfseries{Method}} & \multicolumn{2}{c|}{\bfseries{Peak 
I}} & 
\multicolumn{2}{c|}{\bfseries{Peak II}} & \multicolumn{2}{c|}{\bfseries{Peak 
III}} \\
  & Pos. (eV) & Max. & Pos. (eV) & Max. & Pos. (eV) & Max. \\
  \hline
  4x4x4 & 3.19 &126.03 & 4.19 &74.53 & 5.13 &17.26 \\
  4x4x4 + M1(8NB) & 3.46 &41.69 & 4.23 &63.32 & 5.34 &13.47 \\
  4x4x4 + M2(8NB) & 3.35 &41.44 & 4.10 &64.31 & 5.22 &14.28 \\
  4x4x4 + M3(8NB) & 3.35 &43.17 & 4.11 &64.50 & 5.22 &14.06 \\
  4x4x4 + M1(1NB) & 3.44 &44.02 & 4.21 &60.66 & 5.33 &14.68 \\
  4x4x4 + M2(1NB) & 3.36 &41.90 & 4.12 &62.59 & 5.25 &13.93 \\
  4x4x4 + M3(1NB) & 3.36 &43.50 & 4.13 &62.93 & 5.25 &13.81 \\
  4x4x4 + Ref.\cite{Gillet2013} & 3.23 &38.71 & 3.99 &50.97 &  5.15 
&16.38 \\
  8x8x8 &  3.37 &41.25 & 4.14 &60.74 & 5.24 &13.60 \\
  8x8x8 + M1(8NB) & 3.55 &38.13 & 4.28 &50.99 & 5.25 &14.71 \\
  8x8x8 + M2(8NB) & 3.51 &36.81 & 4.24 &51.58 & 5.19 &15.02 \\
  8x8x8 + M3(8NB) & 3.51 &37.71 & 4.23 &51.56 & 5.19 &14.83 \\
  8x8x8 + M1(1NB) & 3.55 &38.72 & 4.27 &51.00 & 5.21 &14.29 \\
  8x8x8 + M2(1NB) & 3.51 &36.62 & 4.24 &51.05 & 5.19 &14.80 \\
  8x8x8 + M3(1NB) & 3.51 &37.59 & 4.23 &50.98 & 5.18 &14.66 \\
  8x8x8 + Ref.\cite{Gillet2013} & 3.58 &36.49 & 4.21 &48.34 & 5.17 &14.43 
\\
  16x16x16 &  3.50 &38.37 & 4.23 &50.80 & 5.19 &14.57 \\
  \hline
 \end{tabular}
 \caption{Peak position (Pos.) and maximum amplitude (Max.) of the three main 
peaks of the 
absorption spectra of silicon represented in 
Figure~\ref{fig:interpsi}. See the caption of the figure for a complete 
description 
of the
notations.}
 \label{tab:interpsi}

\end{table}
\begin{table}
 \centering
 \begin{tabular}{|c|cc|cc|cc|}
  \hline
  \multicolumn{1}{|c|}{\bfseries{Method}} & \multicolumn{2}{c|}{\bfseries{Peak 
I}} & 
\multicolumn{2}{c|}{\bfseries{Peak II}} & \multicolumn{2}{c|}{\bfseries{Peak 
III}} \\
  & Pos. (eV) & Max. & Pos. (eV) & Max. & Pos. (eV) & Max. \\
  \hline
  4x4x4 & 1.70 &31.67 & 2.62 &104.08 & 4.34 &72.43 
\\
  4x4x4 + M1(8NB) & 1.93 &4.75 & 2.86 &37.60 & 4.52 &35.65 \\
  4x4x4 + M2(8NB) & 1.82 &4.62 & 2.73 &36.39 & 4.37 &33.79 \\
  4x4x4 + M3(8NB) & 1.82 &4.67 & 2.73 &37.17 & 4.36 &33.52 \\
  4x4x4 + M1(1NB) & 1.93 &4.85 & 2.83 &37.79 & 4.48 &31.67 \\
  4x4x4 + M2(1NB) & 1.82 &4.62 & 2.74 &36.08 & 4.39 &33.57 \\
  4x4x4 + M3(1NB) & 1.82 &4.67 & 2.74 &36.85 & 4.39 &33.27 \\
  4x4x4 + Ref.\cite{Gillet2013} & 1.70 &4.53 & 2.60 &33.93 & 4.28 &30.57 
\\
  8x8x8 &  1.82 &4.57 & 2.74 &36.16 & 4.40 &34.45 \\
  8x8x8 + M1(8NB) & 2.04 &2.95 & 3.00 &25.08 & 4.47 &38.08 \\
  8x8x8 + M2(8NB) & 2.00 &2.83 & 2.95 &24.66 & 4.42 &38.14 \\
  8x8x8 + M3(8NB) & 2.00 &2.87 & 2.94 &25.00 & 4.41 &38.12 \\
  8x8x8 + M1(1NB) & 2.03 &3.12 & 2.99 &25.92 & 4.45 &37.36 \\
  8x8x8 + M2(1NB) & 1.98 &2.88 & 2.95 &25.06 & 4.42 &38.02 \\
  8x8x8 + M3(1NB) & 1.98 &2.91 & 2.94 &25.40 & 4.41 &37.77 \\
  8x8x8 + Ref.\cite{Gillet2013} & 1.91 &2.69 & 2.91 &23.32 & 4.38 &35.58 
\\
  16x16x16 & 1.98 &2.95 & 2.93 &25.84 & 4.41 &37.74 \\
  \hline
 \end{tabular}
 \caption{Peak position (Pos.) and maximum amplitude (Max.) of the three main 
peaks of the 
absorption spectra of gallium arsenide represented in 
Figure~\ref{fig:interpgaas}. See the caption of the figure for a complete 
description 
of the
notations.}
 \label{tab:interpgaas}

\end{table}

\begin{table}
 \centering
 \begin{tabular}{|c|cc|cc|cc|}
  \hline
  \multicolumn{1}{|c|}{\bfseries{Method}} & \multicolumn{2}{c|}{\bfseries{Peak 
I}} & 
\multicolumn{2}{c|}{\bfseries{Peak II}} & \multicolumn{2}{c|}{\bfseries{Peak 
III}} \\
  & Pos. (eV) & Max. & Pos. (eV) & Max. & Pos. (eV) & Max. \\
  \hline
  4x4x4 & 11.77 &15.52 & 12.96 &7.15 & 14.22 & 2.02 \\
  4x4x4 + M1(8NB) & 12.67 &15.02 & 14.06 &2.23 & 14.61 &1.49 \\
  4x4x4 + M2(8NB) & 12.24 &13.77 & 13.37 &2.11 & 13.83 &3.37 \\
  4x4x4 + M3(8NB) & 12.08 &14.92 & 13.35 &1.80 & 13.82 &2.61 \\
  4x4x4 + M1(1NB) & 12.12 &15.88 & 13.27 &1.47 & 13.72 &1.81 \\
  4x4x4 + M2(1NB) & 12.10 &16.82 & 13.37 &1.95 & 13.79 &3.07 \\
  4x4x4 + M3(1NB) & 11.91 &17.91 & 13.35 &1.57 & 13.78 &2.34 \\
  4x4x4 + Ref.\cite{Gillet2013} & 11.97 &11.71 & 12.96 &1.32 & 13.43 
&2.21 \\
  8x8x8 & 12.0 &18.4 & 13.35 &1.85 & 13.77 &2.62 \\
  8x8x8 + M1(8NB) & 12.20 &18.00 & 13.88 &3.04 & 14.21 &1.75 \\
  8x8x8 + M2(8NB) & 12.12 &16.52 & 13.56 &3.46 & 14.00 &1.72 \\
  8x8x8 + M3(8NB) & 11.95 &17.80 & 13.54 &3.05 & 14.00 &1.51 \\
  8x8x8 + M1(1NB) & 12.02 &18.06 & 13.52 &2.68 & 13.72 &1.81 \\
  8x8x8 + M2(1NB) & 12.07 &17.16 & 13.54 &3.53 & 13.99 &1.65 \\
  8x8x8 + M3(1NB) & 11.90 &18.31 & 13.51 &3.34 & 13.86 &1.47 \\
  8x8x8 + Ref.\cite{Gillet2013} & 11.98 &18.39 & 13.47 &2.23 & 13.77 
&1.62 \\
  16x16x16  & 11.99 &18.49 & 13.52 &3.22 & 13.99 &1.51 \\
  \hline
 \end{tabular}
 \caption{Peak position (Pos.) and maximum amplitude (Max.) of the three main 
peaks of the 
absorption spectra of lithium fluoride represented in 
Figure~\ref{fig:interplif}. See the caption of the figure for a complete 
description 
of the
notations.}
 \label{tab:interplif}
\end{table}

\begin{figure}
 \subfloat[$4 \times 4 \times 4$ with 1 neighbor]{ \def\svgwidth{7cm}
 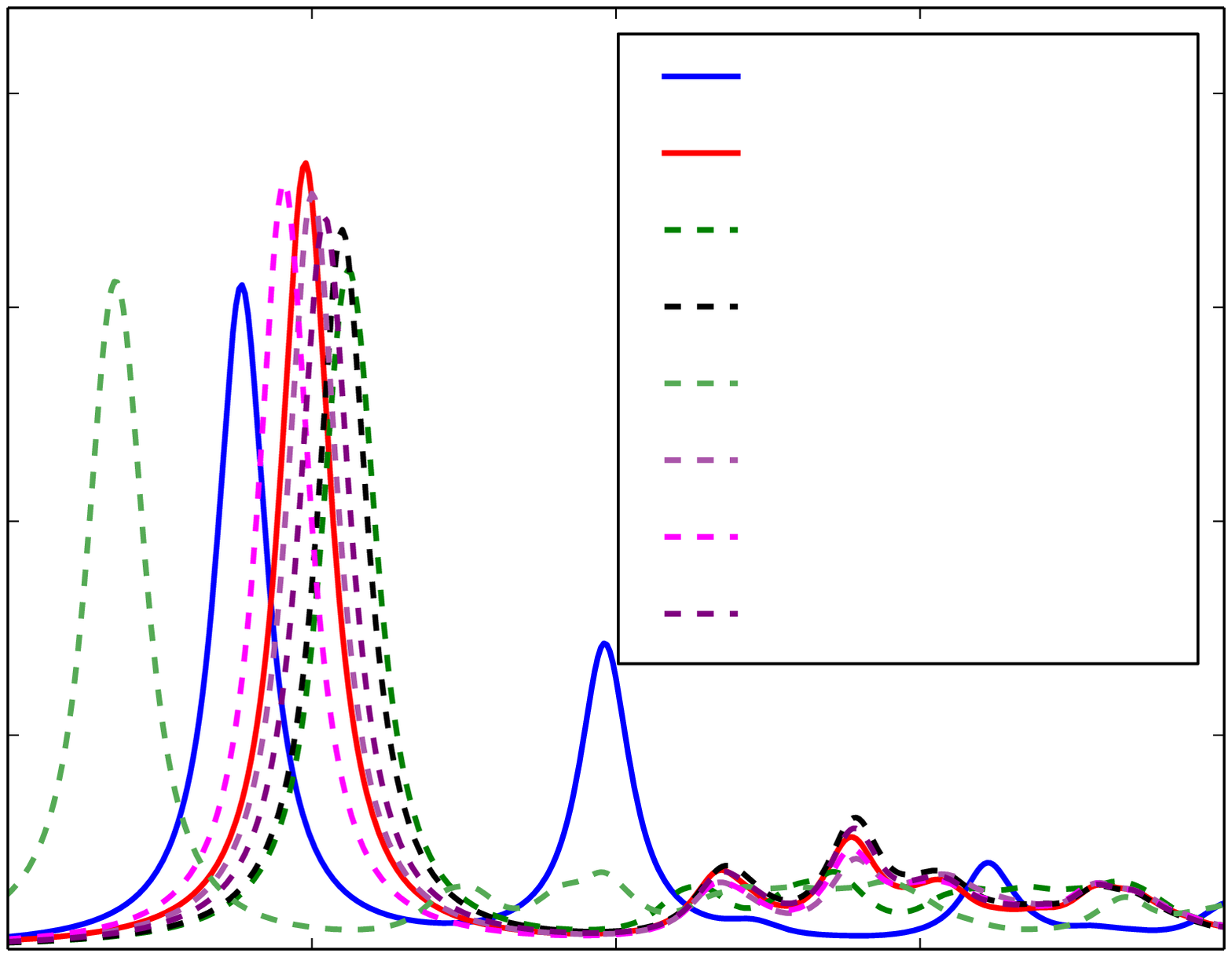} 
 \subfloat[$4 \times 4 \times 4$ with 8 neighbors]{\def\svgwidth{7cm}
 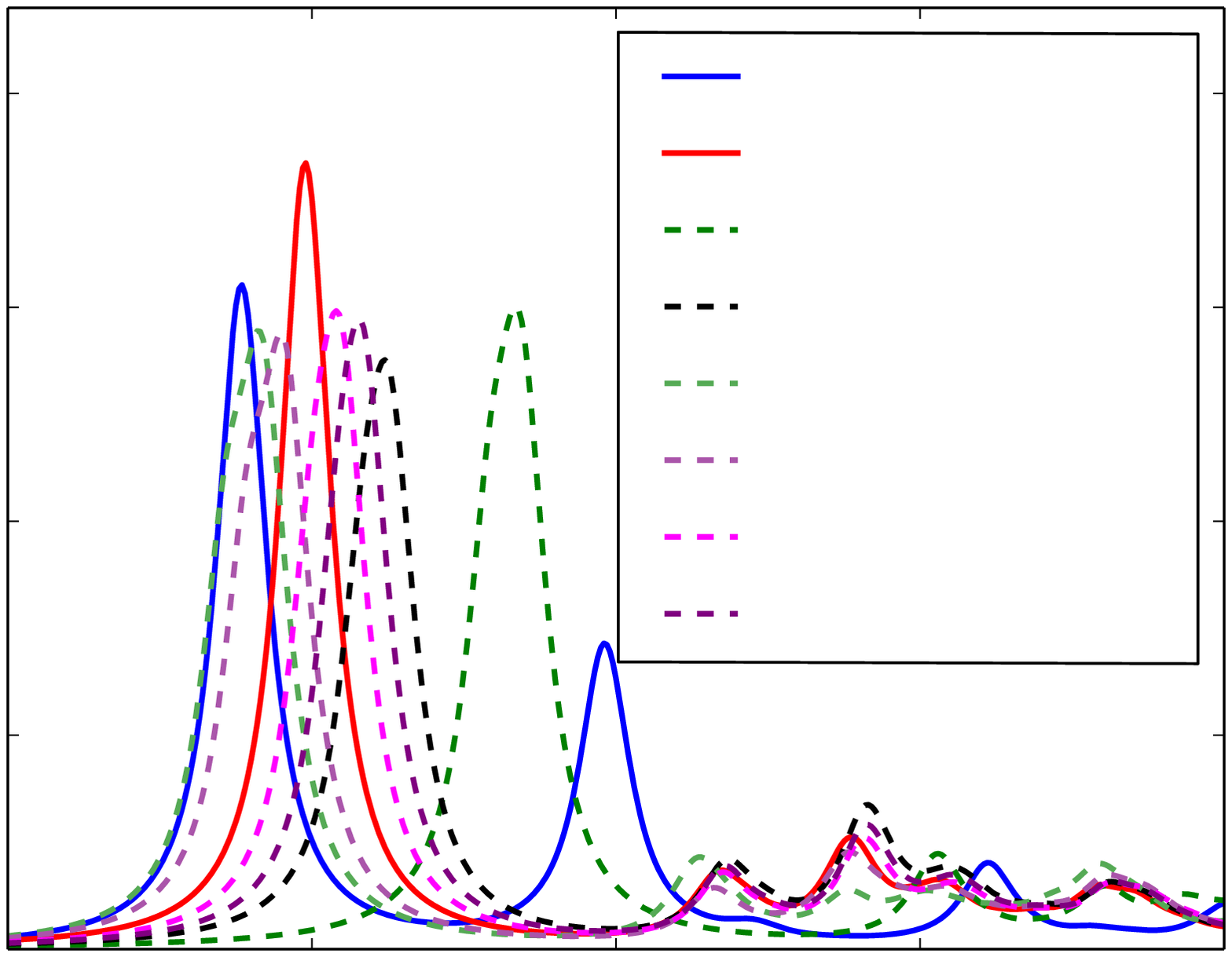}
 \caption{(Color online). Optical absorption spectrum of LiF obtained with 
different values of the width $w$ used for the treatment of the divergence.}
 \label{fig:m3widthlif}
\end{figure}

By comparing the interpolation schemes with 8 neighbors (8NB) and standard BSE
computations done on the dense mesh, we observe that M1~(8NB) tends to shift 
the entire spectrum 
by a small energy and 
the excitonic binding energy is therefore underestimated.
M2~(8NB) and M3~(8NB) give similar results for Si and GaAs 
that are almost on top of the computation done on the dense mesh. 
The case of lithium fluoride is more complicated to interpret. 
In this system, indeed, M1~(8NB) gives inaccurate results for 
the position of the first exciton (0.2 eV of error for a $8\times8\times8$ 
coarse mesh). M2~(8NB) performs better 
than M1~(8NB) although the error in the position of the first peak is still on 
the order of 0.12 eV. M3~(8NB) gives the best results: the excitonic 
binding energy is reproduced with 0.05 eV error and also the behavior at higher 
frequency is correctly reproduced.
It should be stressed, however, that this agreement is somehow fortuitous and 
related to the particular value of the width $w$ used for the treatment of the 
divergence.
Figure~\ref{fig:m3widthlif} shows the optical spectra of LiF  computed with M3 
and different values of $w$. 
Our results indicate that the value of the width used in M3 has a 
significant impact on the position of the first peak of LiF. Therefore some sort 
of convergence study is needed for M3 in order to find values of $w$ 
giving a 
good compromise between accuracy and efficiency.

If we compare the method using one neighbor (1NB) and 
the eight neighbors (8NB), we observe that the original Rohlfing and Louie 
interpolation (1NB)
gives results for GaAs and Si whose quality is comparable to the multilinear 
interpolation and even better results for the special case of LiF. This is 
somewhat puzzling. Our current understanding is as follows.
As already mentioned in the 
previous paragraph, the description of the divergent behavior along the 
diagonal of the Hamiltonian for LiF is extremely important to get a correct 
binding energy. In practice, the number of bands used in  
Eq.~\eqref{eq:completeness} must be truncated and therefore the expansion is not 
exact. Furthermore, we neglect in the expansion possible contributions to 
valence (conduction) states coming from the conduction (valence) manifold in 
Eq.~\eqref{eq:expansion}. This approximation is also used in 
Ref.~\cite{Rohlfing2000}.
Some terms are therefore neglected and they lead to some loss 
of information when building the interpolated matrix element from multiple 
neighbors.

Our results indicate that, although the multilinear interpolation was 
expected to give more accurate results, the practical implementation and 
the numerical approximations tend to favor a ``simple'' 1-neighbor 
interpolation.
This interpolation gives sufficient accuracy at a lower computational cost 
as 
summations over 1 neighbor are cheaper than summations over 8 neighbors.

For the sake of completeness, we have also compared our methods with the 
multiple-shift technique 
introduced in Ref.~\cite{Paier2008,Gillet2013} and used recently in 
Ref.~\cite{Sander2015}. Different coarse grids are obtained by shifting an 
initial homogeneous mesh so that the full set of points forms a 
much denser sampling. An approximate dielectric function is 
then obtained 
by averaging the results obtained on the coarse grids.
As can be seen in 
Fig.~\ref{fig:interpsi}, \ref{fig:interpgaas} and \ref{fig:interplif}, this 
technique tends to smooth the 
spectrum and the amplitude of the peaks is underestimated. As stated in 
Ref.~\cite{Sander2015}, due to the 
localized character of the exciton in LiF, a small number of points in the 
coarse grids is enough to converge the peak position but the correct description 
of the fine details of the spectrum requires more accurate methods. The 
methods developed in the present article are more accurate than this technique 
and are significantly cheaper as they do not require multiple expensive 
calculations of BSE Hamiltonians.

As regards computational efficiency, one should notice that the time 
required to produce an interpolated spectrum for LiF in 
sequential with the above-mentioned parameters is respectively 22200~sec for M1 
(8NB),
120000~sec for M2 (8NB), 3000~sec for M1 (1NB) and 
80000~sec for M2 (1NB).
As a reference, the time needed to compute the matrix elements of the BSE 
Hamiltonian on the coarse mesh is around 15000~sec and 
 around $1\times10^6$ sec for the dense mesh.
To sum up, M1 leads to a gain of two orders of 
magnitude in terms of CPU time while the high-accuracy M2 gives a speedup of 
one order of magnitude. 
The memory required by M1 is of the same order as the one needed for a 
calculation on the coarse mesh whereas M2 is much more memory demanding since 
the  whole dense matrix must be stored.

The technique based on multiple shifts, on the other hand, requires 8 
calculations of a coarse Hamiltonian.
These calculations are independent and can be executed in parallel but the final 
results cannot reach the same frequency resolution as the ones obtained with a 
fast 
interpolation on a dense k-mesh. 

\subsection{Numerical scaling of the interpolation technique}

In order to assess the numerical scaling of our implementation, we have 
performed several benchmarks for silicon with unconverged parameters. A 
cut-off energy of 4 Ha has been used for the wavefunctions and 2 Ha for the 
dielectric function. Only one valence and one conduction band are included in 
the Bethe-Salpeter kernel. This allowed us to increase the number of 
wavevectors 
of the dense grid to more than 100000 wavevectors in the BZ.

\begin{figure*}
 \begin{tabular}{ccc}
  \includegraphics[width=7cm]{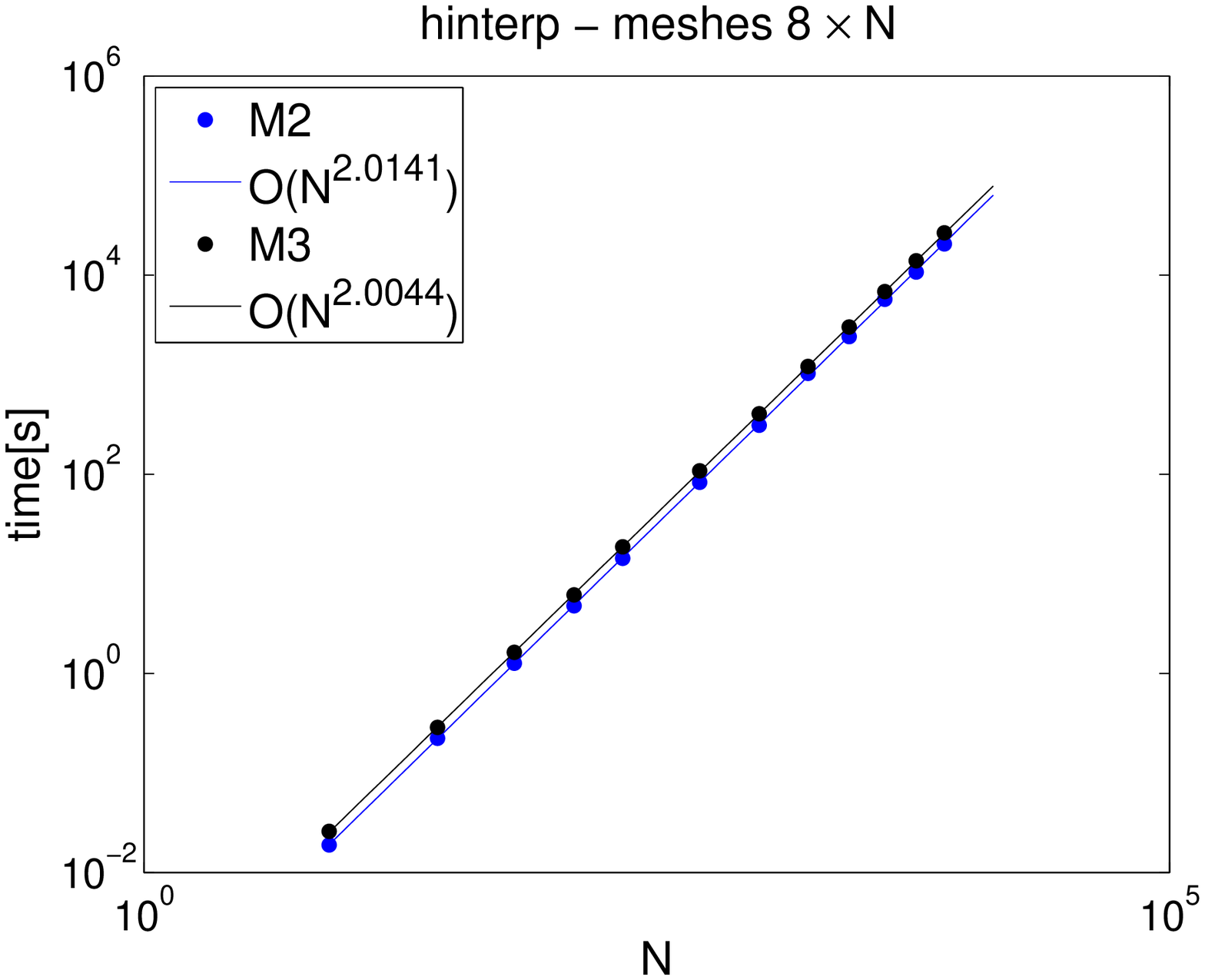} &
  \includegraphics[width=7cm]{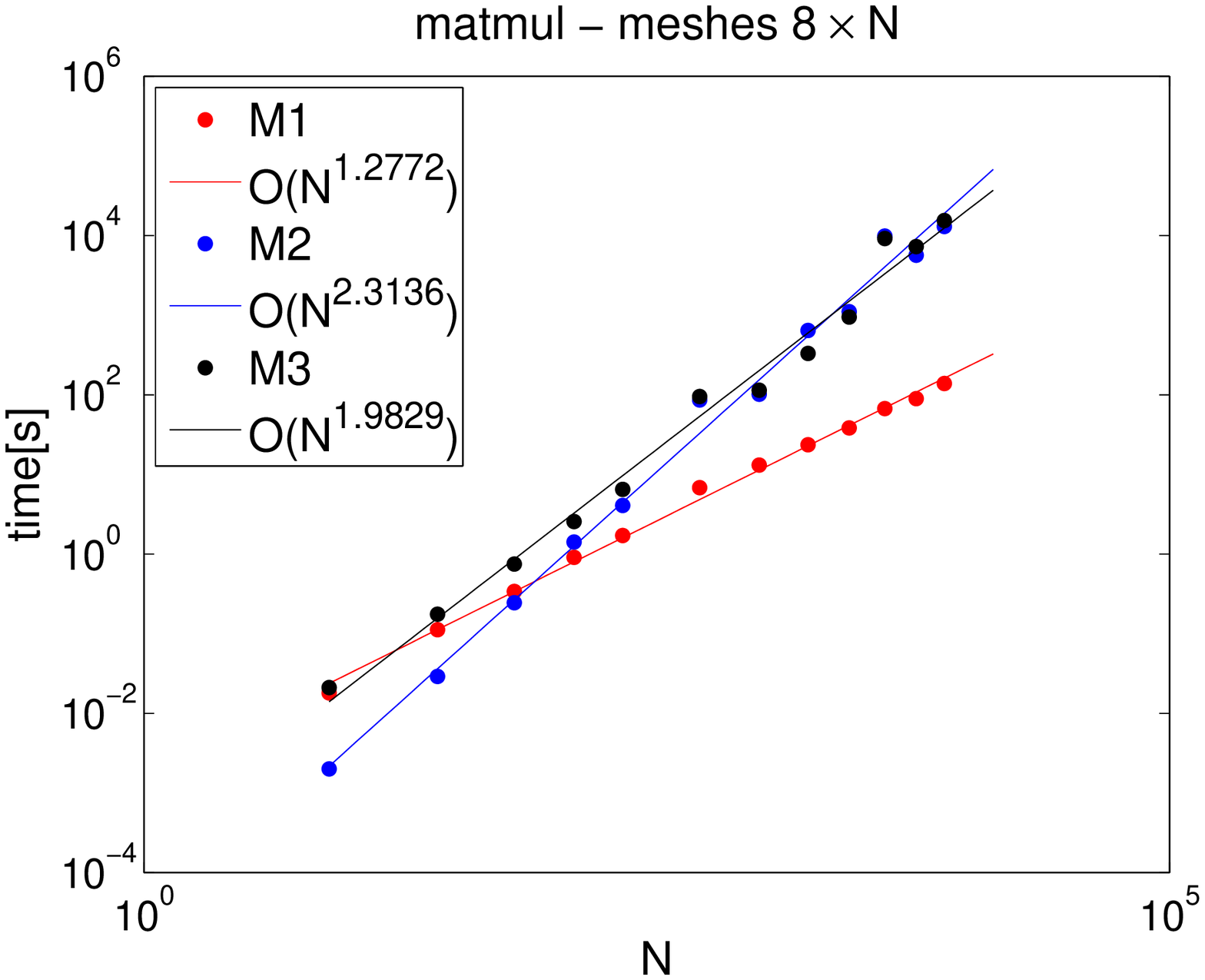} \\
  \includegraphics[width=7cm]{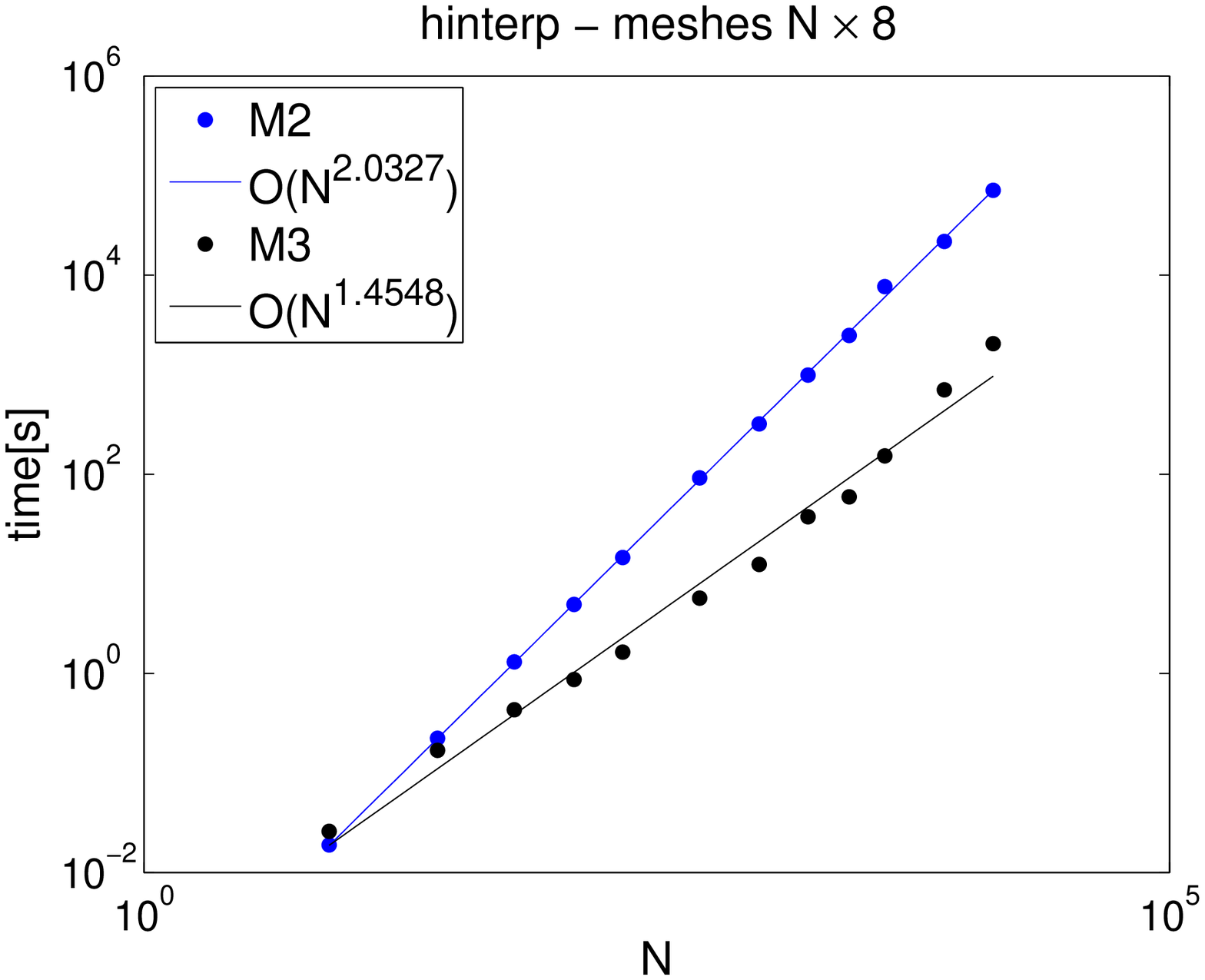} &
  \includegraphics[width=7cm]{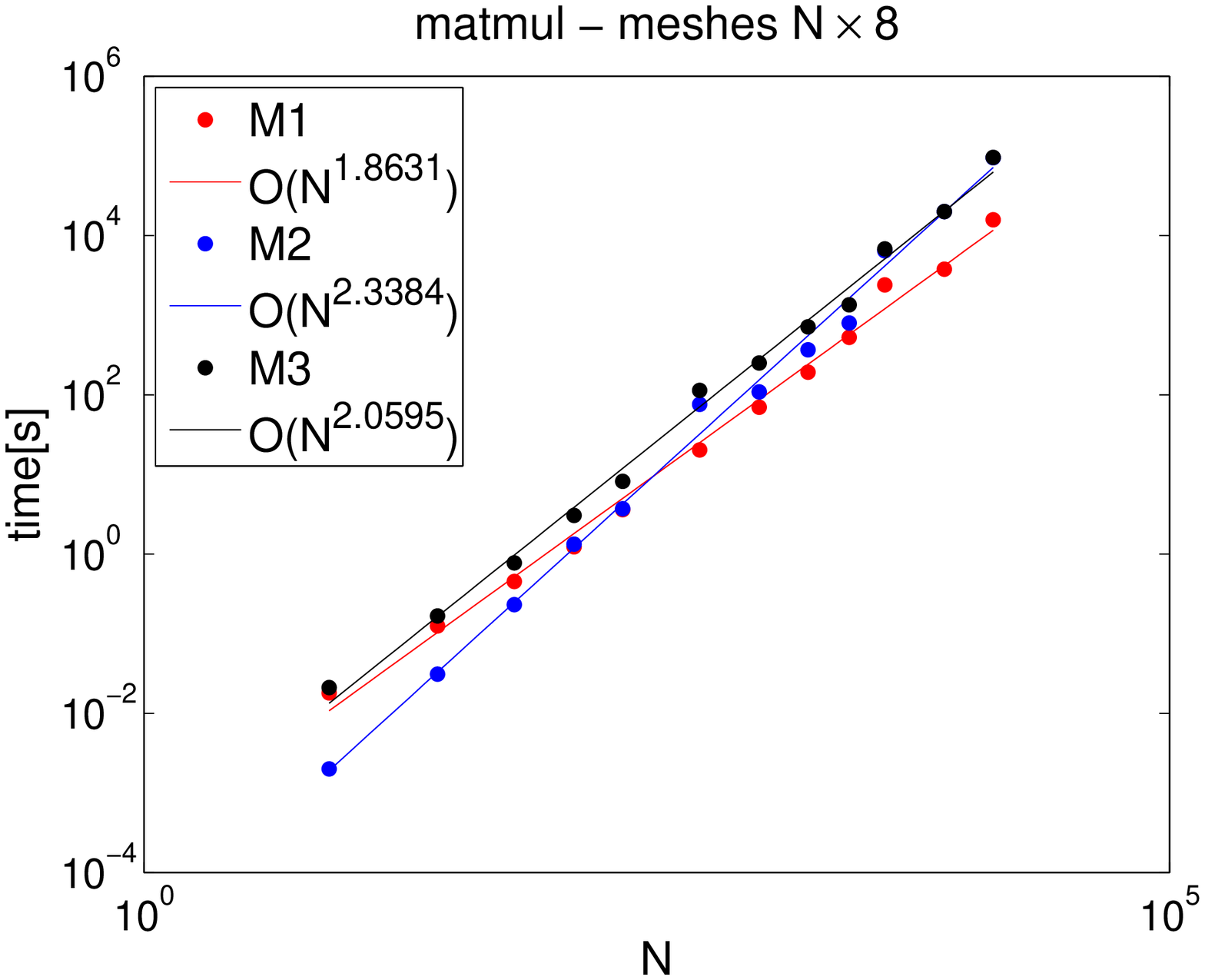} \\
  \includegraphics[width=7cm]{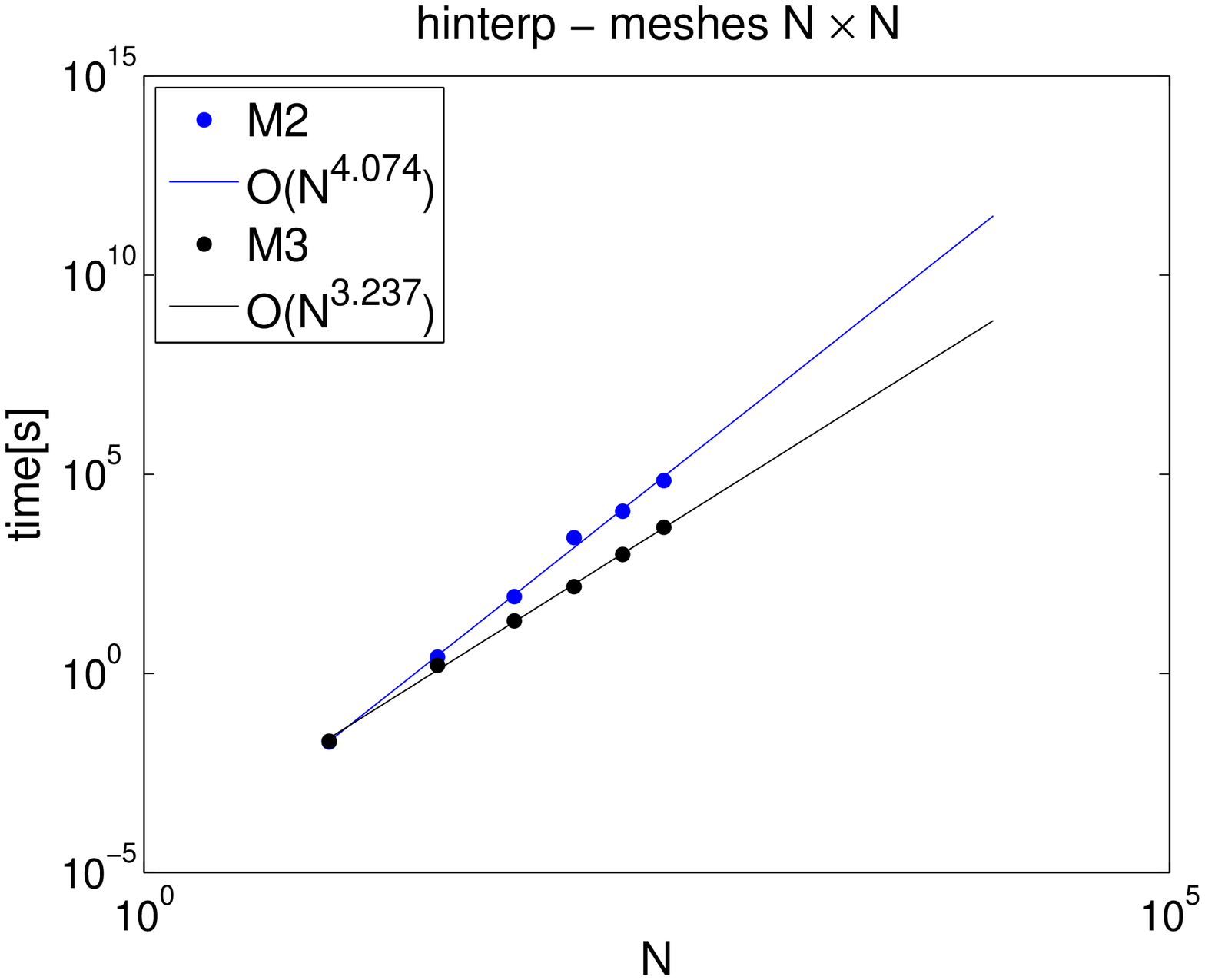} &
  \includegraphics[width=7cm]{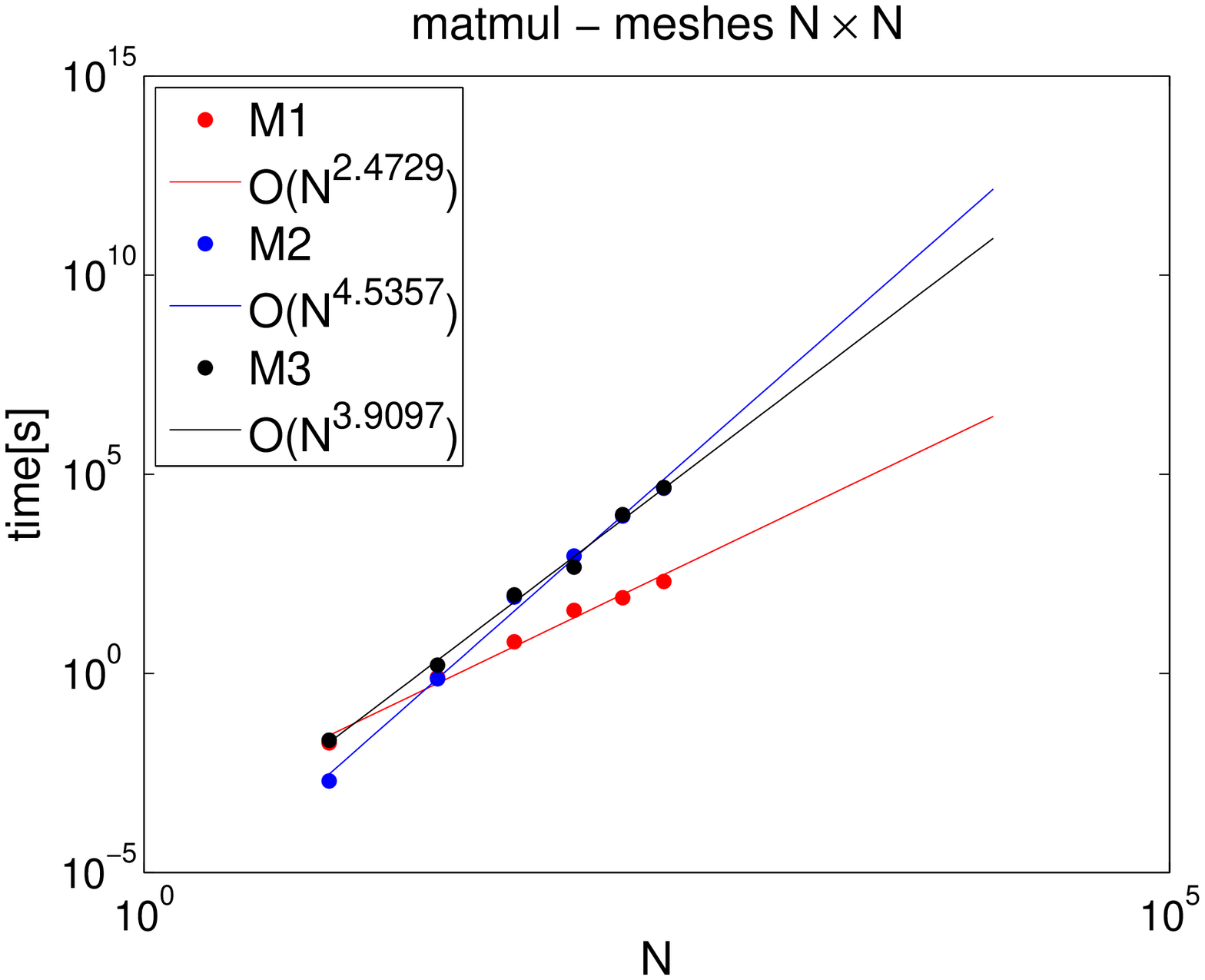} \\
 \end{tabular}
 \caption{Measured scaling for the three interpolation 
methods\label{fig:scalings}}
\end{figure*}

For the three different methods, we have analyzed the time spent in the most 
important routines. 
Different benchmarks have been performed by changing the initial coarse grid as 
well 
as the final dense mesh of $\tilde{N}_k \times N_{div}$ wavevectors. 
The most CPU-critical sections are \texttt{Hinterp} for 
the calculation of the interpolated matrix elements and \texttt{Matmul} for the 
matrix-vector multiplications needed for the Lanczos method.
The theoretical scaling is given in Table~\ref{tab:scalings} while the results 
of the tests are reported in Fig.~\ref{fig:scalings}. 
Several interesting observations on the major trends can be derived from the 
benchmarks.  
If we look at the interpolated matrix-vector product (\texttt{Matmul}), we 
observe that M1 is very efficient as it scales with the square of the size of 
the coarse mesh. On the other hand, both M2 and M3 are less performant. Finally, 
the time spent by M3 in the routine \texttt{Hinterp} (interpolation of matrix 
elements) is much smaller than the one spent by M2 at dense meshes.

\begin{table}
 \centering
  \begin{tabular}{r|cc}
   &\texttt{Matmul}&\texttt{Hinterp} \\
   \hline
   M1 & $\mathcal{O}(\tilde{N}_k^2 + \tilde{N}_k N_{div})$ & - \\
   M2 & $\mathcal{O}(\tilde{N}_k^2 N_{div}^2)$ & $\mathcal{O}(\tilde{N}_k^2 
N_{div}^2)$ \\
   M3 & $\mathcal{O}(\tilde{N}_k N_{div}^2 + \tilde{N}_k^2)$[*] & 
$\mathcal{O}(\tilde{N}_k 
N_{div}^2)$ \\
  \end{tabular}
  \caption{Theoretical scalings of the routines used in the three methods 
described in the text\label{tab:scalings}. [*] Scaling of an optimal 
implementation that takes advantage of sparse matrices. The scaling becomes 
$\mathcal{O}(\tilde{N}_k^2 N_{div}^2)$ if the method is solved with dense 
matrices.}

\end{table}

\subsection{Comparison with the Kammerlander double-grid technique}

In this section, we compare our method with the technique proposed 
by Kammerlander in Ref.~\cite{Kammerlander2012}.
In that approach, the polarizability $L(\omega)$ is expressed in the transition 
space according to 
\begin{align}
 L_{vc\boldsymbol{k},v'c'\boldsymbol{k}'}(\omega) = 
\sum_{v''c''\boldsymbol{k}''} (1-L^0(\omega) 
K)^{-1}_{vc\boldsymbol{k},v''c''\boldsymbol{k}''} 
L^0_{v''c''\boldsymbol{k}'',v'c'\boldsymbol{k}'}(\omega),
\end{align}
where the RPA polarizability $L^0(\omega)$ is given by
\begin{align}
 L^0_{vc\boldsymbol{k},v'c'\boldsymbol{k}'}(\omega) = 
\frac{f_{c\boldsymbol{k}}-f_{v\boldsymbol{k}}}{\varepsilon_{c\boldsymbol{k}} - 
\varepsilon_{v\boldsymbol{k}} - \omega - i \eta} \delta_{cc'} \delta_{vv'} 
\delta_{\boldsymbol{k},\boldsymbol{k}'}.
\end{align}

In order to avoid the direct inversion of the large matrix, an iterative scheme 
is used for the computation of 
\begin{align}
 L_{vc\boldsymbol{k},v'c'\boldsymbol{k}'}(\omega) = \sum_{m} 
\sum_{v''c''\boldsymbol{k}''} \left[ 
L^0(\omega) 
K
\right]^m_{vc\boldsymbol{k},v''c''\boldsymbol{k}''}  
L^0_{v''c''\boldsymbol{k}'',v'c'\boldsymbol{k}'}(\omega) \label{eqiterK}.
\end{align}

The BSE is solved for every frequency in an iterative way and a double grid 
technique is used to reduce the number of k-points for which the kernel must be  
computed explicitly. The RPA polarizability is averaged on a 
dense mesh yielding
\begin{align}
 L^0_{vc\boldsymbol{k},v'c'\boldsymbol{k}'}(\omega) = \frac{1}{N_{nb}} 
\sum_{\bar{\boldsymbol{k}} \in N(\boldsymbol{k})}
\frac{f_{c\bar{\boldsymbol{k}}}-f_{v\bar{\boldsymbol{k}}}}{\varepsilon_{
c\bar{\boldsymbol{k}}} - 
\varepsilon_{v\bar{\boldsymbol{k}}} - \omega - i \eta} 
\delta_{cc'} 
\delta_{vv'} 
\delta_{\boldsymbol{k},\boldsymbol{k}'},
\end{align}
where the $\bar{\boldsymbol{k}}$ are taken from the set $N(\boldsymbol{k})$ of 
$N_{nb}$ dense points located around $\boldsymbol{k}$.

Finally the averaged values are used in the iterative BSE solver [see Eq. 
\eqref{eqiterK}]. This approach has the advantage that the 
wavefunctions on the dense mesh are not needed but the 
divergence is not accurately reproduced.
The scaling of the Kammerlander technique is linear with the number of 
frequencies and quadratic with the number of points in the coarse 
mesh. On the contrary, our technique is able to describe the frequency 
dependence with a computational cost that does not depend on the number of 
frequency points, since, as discussed in Section~\ref{sec2}, 
$\varepsilon_M(\omega)$ is evaluated 
with 
Eq.~\eqref{eq:fraccont} whose cost is negligible.

\subsection{Wavefunction interpolation}

In this article, we assume that the entire set of wavefunctions on the dense 
set of points is available. For the systems investigated in this study, the calculation of 
wavefunctions starting from an already converged density is not the most 
computationally demanding part. Moreover, only wavefunctions in the transition 
basis set are required, that is a small fraction of the set of bands required 
for the screening, for example.

However, for some more complex systems, one might take advantage of 
interpolation techniques to obtain the wavefunctions on denser meshes. In 
the work of Kammerlander~\cite{Kammerlander2012} presented in the previous 
section, Wannier functions were used to obtain 
eigenenergies on these dense meshes. Recently, Gilmore \textit{et 
al.}~\cite{Gilmore2015} have used optimized basis functions described in the work 
of Shirley~\cite{Shirley1996} to compute wavefunctions on a dense mesh.
These different techniques could be easily interfaced with our technique to 
compute the overlap matrix elements, that can afterwards be used in the 
interpolation of the BSE Hamiltonian.

\section{Conclusions}

We have presented a fast and memory-efficient technique 
that combines the interpolation of the Bethe-Salpeter matrix elements with the 
Lanczos algorithm.
The treatment of the matrix elements is similar in spirit to the Rohlfing 
and Louie approach but we avoid the storage and the 
diagonalization of large matrices.
Three possible approaches for the treatment of the Coulomb singularity have been 
presented and discussed in detail.

The effectiveness of the method has been analyzed through calculations of 
optical spectral in Si, GaAs and LiF.
According to our tests, the multilinear interpolation of the wavefunctions does 
not perform better than simple constant interpolation, already proposed by 
Rohlfing and Louie (although used by them only for the set up of the 
Hamiltonian on the dense mesh).

In conclusion, we suggest using Method 1 for a quick qualitative analysis of 
optical spectra e.g. for a high-throughput screening to rapidly identify 
possible candidates.
Method 3 with the on-the-fly interpolation is the recommended approach  
for BSE calculations requiring dense k-meshes since
it is significantly faster than M2 and the Coulomb divergence is 
taken into account.
The downside is that one has to check the convergence with the width $w$, but 
we believe this is a small price to pay, especially when compared with the 
significant speedup that can be achieved.

The algorithmic improvements presented in this work will facilitate BSE 
calculations in complex systems and will also significantly ease the ab 
initio study of piezoreflectance, thermoreflectance and Raman intensities in 
systems with 
excitonic effects.

\section{Acknowledgments}
Y.G. and M.G. wish to acknowledge the financial support of the Fonds National 
de la Recherche Scientifique (FNRS, Belgium).
The authors would like to thank Yann Pouillon and Jean-Michel Beuken for their 
valuable technical support and help with the test and build system of ABINIT.

Computational resources have been provided by the supercomputing facilities of
the Universit\'e catholique de Louvain (CISM/UCL) and the Consortium des
Equipements de Calcul Intensif en F\'ed\'eration Wallonie Bruxelles (CECI) funded by
the Fonds de la Recherche Scientifique de Belgique (FRS-FNRS) under Grant No. 
2.5020.11.
This work was also supported by the FRS-FNRS Belgium through PDR Grant T.0238.13 - AIXPHO.

\section{References}
\bibliographystyle{elsarticle-num}
\bibliography{interp}

\end{document}